\documentclass[preprint,showpacs,preprintnumbers,superscriptaddress,amsmath,amssymb]{revtex4}

\usepackage{epsfig}
\usepackage{dcolumn}
\usepackage{bm}

\begin{document}


\title{Testing for non-Gaussianity of the cosmic microwave background in harmonic space: an empirical
process approach}

\author{Frode K. Hansen}
\email{frodekh@roma2.infn.it}
\affiliation{Dipartimento di Fisica, Universit\`a di Roma `Tor Vergata', Via della Ricerca Scientifica 1, I-00133 Roma, Italy}
\author{Domenico Marinucci}
\email{domenico.marinucci@uniroma1.it}
\affiliation{Dipartimento di Studi Geoeconomici e Statistici, Universit\`a di Roma 'La Sapienza', Via del Castro Laurenziano 9, I-00161 Rome, Italy}
\author{Paolo Natoli}
\affiliation{Dipartimento di Fisica, Universit\`a di Roma `Tor Vergata', Via della Ricerca Scientifica 1, I-00133 Roma, Italy}
\author{Nicola Vittorio}
\affiliation{Dipartimento di Fisica, Universit\`a di Roma `Tor Vergata', Via della Ricerca Scientifica 1, I-00133 Roma, Italy}

\date{\today}

\begin{abstract}
We present a new, model-independent approach for measuring non-Gaussianity of the Cosmic Microwave Background (CMB) anisotropy pattern. Our approach is based on the empirical distribution function of the normalized spherical harmonic expansion coefficients $a_{\ell m}$ of a nearly full-sky CMB map, like the ones expected from forthcoming satellite experiments. Using a set of Kolmogorov-Smirnov type tests, we check for Gaussianity and independency of the $a_{\ell m}$. We test the method on two non-Gaussian toy-models of the CMB, one generated in spherical harmonic space and one in pixel (real) space. We also provide some rigorous results, possibly of independent interest, on the exact distribution of the spherical harmonic coefficients normalized by an estimated angular power spectrum. 
 
\end{abstract}

\maketitle

\section{Introduction}
\label{sect:intro}

Temperature fluctuations in the CMB are an invaluable tool to constraint cosmological models and the process of structure formation in the universe. According to the standard version of the most popular theory of the very early universe, the so-called theory of inflation, the density fluctuations at primordial epochs should be Gaussian or very close to Gaussian distributed (see the reviews \cite{narlikar,watson}). This implies that the temperature fluctuations in the CMB as observed today should also be close to Gaussian, because they are related by linear theory to fluctuations in the early universe. However, the details of the inflationary scenario are still rather unclear. Newer, more sophisticated, versions of inflation predict small deviations from Gaussianity \cite{nongi1,nongi2,nongi3,nongi4,nongi5,nongi6}. Detecting this non-Gaussianity in the CMB would thus be important for understanding the physics of the very early universe. There are also other possible mechanisms for the creation of  non-Gaussianity in the CMB. If the universe has undergone a phase transition at early times, this could have given rise to topological defects (See Ref.\cite{strings} for a review). These defects would show up as non-Gaussian features in the CMB temperature fluctuation field.\\

With the high resolution
data from the satellite missions MAP\footnote{http://map.gsfc.nasa.gov/} and Planck Surveyor\footnote{http://astro.estec.esa.nl/SA-general/Projects/Planck/} one could be able to detect possible deviations from non-Gaussianity in the CMB. If this is indeed detected it would have a big impact on our understanding of the physics of the early universe. For these
reasons, the search for procedures to test for Gaussianity in high resolution data has recently
drawn an enormous amount of attention in the CMB literature. A number of
methods were proposed, many based upon topological properties of spherical
Gaussian fields: the behavior of Minkowski functionals \cite{novikov,gott}, temperature correlation functions \cite{eriksen}, the peak to peak correlation function \cite{heavens}, skewness and kurtosis of the temperature field \cite{vittorio} and local curvature properties of
Gaussian and chi-squared fields \cite{dore}. Other works have focussed on harmonic space
approaches: analysis of the bispectrum and its normalized version \cite{phillips,komatsu} and 
the bispectrum in the
flat sky approximation \cite{win}. The explicit form of the trispectrum for CMB data was derived in \cite{hu,kunz}.
Applications to COBE, Maxima and Boomerang data have also drawn enormous
attention and raised wide debate \cite{boom,polenta,cobeng1,cobeng2}.\\

Another reason to look for non-Gaussianities in CMB data, is to detect effects of systematic errors in the CMB map. One of these systematic effects can be stripes from the $1/f$ noise in the detectors which have not been properly removed. Another systematics which could induce non-physical non-Gaussianities in the map could be the effect of straylight contamination from the galaxy or distortions of the main beam caused by the optics of the telescope. Finally astrophysical foregrounds like synchrotron emission, thermal dust emission or free-free emission from the galaxy and other extra galactic sources could be wrongly interpreted as a physical non-Gaussian signal. It is important that one uses data from simulated experiments to check whether these systematic effects induce non-physical non-Gaussian signatures in the CMB map.\\

Our purpose here is to propose a new procedure to detect non-Gaussianity in
harmonic space. More precisely, let $T(\theta ,\varphi )$ denote the CMB
fluctuations field, which we assume, as always, to be homogeneous and
isotropic, for $0<\theta \leq \pi ,$ $0<\varphi \leq 2\pi $. Assuming that $%
T(\theta ,\varphi )$ has zero mean and finite second moments, it is
well-known that the following spectral representation holds: 
\begin{equation*}
T(\theta ,\varphi )=\sum_{\ell=1}^{\infty }\sum_{m=-\ell}^{\ell}a_{\ell m}Y_{\ell m}(\theta
,\varphi )\text{ ,}
\end{equation*}%
where $Y_{\ell m}(\theta ,\varphi )$ denotes the spherical harmonics. The random
coefficients (amplitudes) $\left\{ a_{\ell m}\right\} $ have zero-mean with
variance $\langle|a_{\ell m}|^{2}\rangle=C_{\ell}$. They are uncorrelated over $\ell$ and $|m|$: $\langle a_{\ell m}a_{\ell^{\prime }m^{\prime }}^{\ast }\rangle=C_{\ell}\delta
_{\ell}^{\ell^{\prime }}\delta _{m}^{m^{\prime }}$, and $a_{\ell,m}=a_{\ell,-m}^{%
\ast }$. The sequence $\left\{ C_{\ell}\right\} $ denotes the angular power
spectrum of the random field and the asterisk complex conjugation.
Furthermore, if $T(\theta ,\varphi )$ is Gaussian, the $\left\{
a_{\ell m}\right\} $ have a complex Gaussian distribution. Upon observing $%
T(\theta ,\varphi )$ on the full sky, the random coefficients can be obtained through the
inversion formula%
\begin{equation}
a_{\ell m}=\int_{-\pi }^{\pi }\int_{0}^{\pi }T(\theta ,\varphi )Y_{\ell m}^{\ast
}(\theta ,\varphi )\sin \theta d\theta d\varphi \text{ , }m=0,\pm 1,...,\pm l%
\text{ },\text{ }l=1,2,...\text{ .}  \label{alm}
\end{equation}%
Our purpose is to study the empirical distribution function for the $\left\{
a_{\ell m}\right\} ,$ and to use these results to implement tests for
non-Gaussianity in harmonic space. We shall assume that the angular power
spectrum is unknown, and the sequence $\left\{ C_{\ell}\right\} $ estimated
from the data; as we show below, this has a nonnegligible effect on the
behavior of the test, no matter how good is the resolution of the experiment%
$.$ The plan of the paper is as follows: the procedure we advocate is
described in Sections \ref{sect:univar} and \ref{sect:multivar}; Section \ref{sect:empres} presents some empirical results,
whereas Section \ref{sect:disc} is devoted to discussion and to directions for further
work. Some mathematical results are collected in the Appendix.

\section{The univariate empirical process}
\label{sect:univar}

To motivate our procedure, we start from the (unrealistic) assumption that
the sequence of coefficients in the angular power spectrum of $T(\theta
,\varphi ),$ i.e. $\left\{ C_{\ell}\right\} _{\ell=1,2,..L},$ is known; here, $L$
is the highest observable multipole, which depends upon the resolution and noise level of
the experiment (for instance, $L\sim 2000$ for Planck)$.$ Now recall that,
under Gaussianity, $|a_{\ell 0}|^{2}/C_{\ell}$ and $2|a_{\ell m}|^{2}/C_{\ell}$ are
standard chi-square variates, with one and two degrees of freedom,
respectively; furthermore, they are independent and identically distributed $%
(i.i.d.)$ over all $\ell$. It is clearly computationally convenient to
introduce a transformation, in order to work with random variables that have
an uniform distribution on $[0,1]$. This can be achieved as follows. Write $%
\Phi _{i}(x),$ $i=1,2,$ for the cumulative distribution function of a
chi-square with $i$ degrees of freedom; recall that%
\begin{equation*}
\Phi _{1}(x)=\int_{0}^{x}\frac{1}{\sqrt{2\pi u}}\exp (-u/2)du\text{ ,}
\end{equation*}%
\begin{equation*}
\Phi _{2}(x)=\int_{0}^{x}\frac{1}{2}\exp (-u/2)du=1-\exp (-x/2)\text{ , }%
\Phi _{2}^{-1}(\alpha )=-2\log (1-\alpha )\ .
\end{equation*}%
Now introduce the Smirnov transformation (see for instance Ref.\cite{shorac}) 
\begin{equation}
u_{\ell 0}=\Phi _{1}\left(\frac{|a_{\ell 0}|^{2}}{C_{\ell}}\right)\text{ , }u_{\ell m}=\Phi _{2}\left(\frac{%
2|a_{\ell m}|^{2}}{C_{\ell}}\right)\text{ , }m=1,2,...,l,\text{ }l=1,2,...L\text{ .}
\label{smirn}
\end{equation}%
It is immediate to see that the random variables of the triangular array $%
\left\{ u_{\ell m}\right\} $ are $i.i.d.$ with a uniform distribution in $[0,1],$
i.e. 
\begin{eqnarray*}
P\left[ u_{\ell m}\leq \alpha \right] &=&P\left[\Phi _{2}\left(\frac{2|a_{\ell m}|^{2}}{C_{\ell}}%
\right)\leq \alpha \right] \\
&=&P\left[\frac{2|a_{\ell m}|^{2}}{C_{\ell}}\leq \Phi _{2}^{-1}(\alpha )\right]=\Phi _{2}\left[\Phi
_{2}^{-1}(\alpha )\right]=\alpha \text{ },
\end{eqnarray*}%
and likewise for $u_{\ell 0},$ because $\Phi _{1},\Phi _{2}$ are strictly
increasing and hence invertible. For $0\leq \alpha \leq 1,$ we can hence
define the \emph{empirical distribution function} 
\begin{equation*}
F_{\ell}(\alpha )=\frac{1}{\ell+1}\sum_{m=0}^{\ell}\underline{\mathbf{1}}(u_{\ell m}\leq \alpha )\text{ ,}
\end{equation*}%
$\underline{\mathbf{1}}(x)$ denoting the indicator function, i.e the function that takes value
unity if $u_{\ell m}$ smaller or equal than$\ \alpha ,$ zero otherwise; $%
F_{\ell}(\alpha )$ evaluates the proportion of observed $u_{\ell m}$ which is
below a certain value, and thus provides a sample analogue of $P\left(
u_{\ell m}\leq \alpha \right) .$ Then, if the distribution of the $u_{\ell m}$ is
uniform indeed, that is, if the $a_{\ell m}$ are actually Gaussian, the
Glivenko-Cantelli Theorem \cite{shorac,dudley}
ensures that, with probability one, $F_{\ell}(\alpha )\rightarrow \alpha ,$
uniformly in $[0,1].$ This can be viewed as a most intuitive conclusion, as
it states that the relative frequency with which a certain event occurs
converges to the probability of the same event, as the number of experiments
grows. It is therefore natural to look at the distance between $F_{\ell}(\alpha
)$ and $\alpha $ for a test of the assumption that the $a_{\ell m}$ are
Gaussian. More precisely, a centered and rescaled version of $F_{\ell}(\alpha )$
provides the (sequence of) \emph{empirical processes}%
\begin{equation*}
G_{\ell}(\alpha )=\sqrt{\ell+1}\left\{ F_{\ell}(\alpha )-\alpha \right\} ,\text{ }%
\alpha \in \lbrack 0,1]\text{ .}
\end{equation*}%
The idea is that, if the $a_{\ell m}$ are not Gaussian, then $F_{\ell}(\alpha
)\rightarrow F^{\prime }(\alpha ),$ for some $F^{\prime }(\alpha )$ such
that $\sup |F^{\prime }(\alpha )-\alpha |>0,$ and hence $G_{\ell}(\alpha )$
will exhibit very high values as a distinctive feature of non-Gaussianity,
at least over some $\alpha $. On the other hand, if the $a_{\ell m}$ are
Gaussian, $F_{\ell}(\alpha )-\alpha $ converges to zero for all $\alpha $,
whereas $G_{\ell}(\alpha )$ converges to a well-known process, the Brownian
bridge \cite{bill,shorac}, the $\sup $ of which has a tabulated distribution which can be used
to derive threshold values.

An apparent deviation from Gaussianity over some $\ell$, however, must be
careful assessed; indeed, in the Gaussian case, the single $G_{\ell}(x)$ are
independently distributed; hence, if we run a test at the 95\% confidence
level, we should expect approximately $5\%$ of \ the multipoles to provide
results above the threshold level. To avoid spurious detections, our aim is
to combine rigorously the information over all different multipoles $\ell$ into
a single statistic. We shall hence focus on the partial sum process 
\begin{equation*}
K_{L}(r,\alpha )=\frac{1}{\sqrt{L}}\sum_{\ell=1}^{[Lr]}G_{\ell}(\alpha )\ ,\text{ }%
\alpha \in \lbrack 0,1]\text{ , }r\in \lbrack 0,1]\text{ ,}
\end{equation*}%
which has not been considered so far in the literature. The idea is to
evaluate what random fluctuations we can expect over the multipoles, if the
underlying field is Gaussian. Note that, if there is any departure from
Gaussianity$,$ then the behavior of $K_{L}(r,\alpha )$ should also detect
the location of the non-Gaussianity in harmonic space.

It is immediate to see that 
\begin{equation*}
\langle K_{L}(r,\alpha )\rangle =0\text{ ,}
\end{equation*}%
whereas simple calculations show also that, 
\begin{equation}
\lim_{L\rightarrow \infty }\langle K_{L}(r_{1},\alpha _{1})K_{L}(r_{2},\alpha
_{2})\rangle =\min (r_{1},r_{2})\left\{ \min (\alpha _{1},\alpha _{2})(1-\max
(\alpha _{1},\alpha _{2}))\right\} \text{ }.  \label{paris}
\end{equation}%
With more effort it is actually possible to establish a much stronger
result, i.e. functional convergence in the distribution; more precisely \cite{mar}, as $L\rightarrow \infty $ 
\begin{equation*}
K_{L}(r,\alpha )\Rightarrow K(r,\alpha )\text{ ,}
\end{equation*}%
where $K(r,\alpha )$ (the so-called Kiefer-Muller process) is a Gaussian
zero mean random field on $[0,1]\times[0,1]$ ($\alpha$ and $r$ in $[0,1]$) with covariance function given in (\ref%
{paris}). Here, $\Rightarrow $ means weak convergence in a functional sense
\cite{dudley}; this is a much stronger notion of convergence than simply
requiring that the distribution of $K_{L}(r,\alpha )$ is asymptotically the
same as the distribution of $K(r,\alpha ),$ for all fixed $r$ and $\alpha .$
The difference can be explained as follows: for statistical inference, we
must actually consider some functional of $K_{L}(r,\alpha ),$ for instance
the $\sup $ in a Kolmogorov-Smirnov type test. Now convergence in
the distribution for every fixed point does not entail convergence of continuous
functionals such as the $\sup ,$ whereas this is granted by the stronger
notion of convergence we are exploiting here.\newline
We are now in a position to relax the restrictive assumption that the
angular power spectrum is known, to consider the more realistic case where
the latter is estimated from the data. We take 
\begin{equation*}
\widehat{C}_{\ell}=\frac{1}{2\ell+1}\sum_{m=-\ell}^{\ell}|a_{\ell m}|^{2},
\end{equation*}%
so that a Smirnov-type transformation analogous to (\ref{smirn}) now reads 
\begin{equation*}
\widehat{u}_{\ell 0}=\Phi _{1}\left(\frac{|a_{\ell 0}|^{2}}{\widehat{C}_{\ell}}\right)\text{ , }%
\widehat{u}_{\ell m}=\Phi _{2}\left(\frac{2|a_{\ell m}|^{2}}{\widehat{C}_{\ell}}\right)\text{ , }%
m=0,1,2,...,l,\text{ }l=1,2,...L\text{ .}
\end{equation*}%
Likewise, we have an estimated empirical distribution function 
\begin{equation*}
\widehat{F}_{\ell}(\alpha )=\frac{1}{\ell+1}\sum_{m=0}^{\ell}\underline{\mathbf{1}}(\widehat{u}_{\ell m}\leq
\alpha )\text{ ,}
\end{equation*}%
and the empirical process with estimated parameters 
\begin{eqnarray*}
\widehat{G}_{\ell}(\alpha ) &=&\sqrt{\ell+1}\left\{ \widehat{F}_{\ell}(\alpha
)-\alpha \right\} \\
\widehat{K}_{L}(\alpha ,r) &=&\frac{1}{\sqrt{L}}\sum_{\ell=1}^{[Lr]}\widehat{G}%
_{\ell}(\alpha )\text{ , }0\leq \alpha \leq 1\text{ , }0\leq r\leq 1\text{ .}
\end{eqnarray*}%
It is important to stress that, due to the presence of estimated parameters,
the normalized random coefficients $\widehat{u}_{\ell m}$ are no longer
independent as $m$ varies, for a fixed $\ell$; on the other hand, independence
across different multipoles $\ell$ is maintained.

Because of course $\widehat{C}_{\ell}$ converges to $C_{\ell}$ as $\ell$ grows, it
might be conjectured that the effect of using estimated parameters becomes
asymptotically negligible, for $L$ large. In fact, we can show that this is
not the case. It is shown in the Appendix that we have,\textbf{\ }as $%
l\rightarrow \infty $%
\begin{eqnarray}
&\langle &\underline{\mathbf{1}}\left( \widehat{u}_{\ell m}\leq \alpha \right) \rangle =\alpha +\frac{b(\alpha )}{\ell}%
+o\left(\frac{1}{\ell}\right)\text{ ,}  \label{drago1} \\
&\langle &\underline{\mathbf{1}}\left( \widehat{u}_{\ell 0}\leq \alpha \right) \rangle =\alpha +O\left(\frac{1}{\ell}\right)\text{
,}  \label{drago2}
\end{eqnarray}%
where%
\begin{equation}
b(\alpha )=(1-\alpha )\log (1-\alpha )+\frac{1}{2}(1-\alpha )\log
^{2}(1-\alpha )\text{ .}  \label{biadef}
\end{equation}%
Here $o\left(\frac{1}{\ell}\right)$ means that the remaining terms go to zero faster than $\left(\frac{1}{\ell}\right)$.
Also, as $L\rightarrow \infty ,$%
\begin{equation}
\lim_{L\rightarrow \infty }\langle \widehat{K}_{L}(\alpha ,r)\rangle =2\sqrt{r}b(\alpha )%
\text{ .}  \label{bias}
\end{equation}%
Equation (\ref{bias}) shows how failing to take into account the presence of
estimated parameters may results in unwarranted conclusions: the limiting
field entails a non-vanishing bias and hence needs further centering before
reliable inference can take place. In view of (\ref{drago1})-(\ref{drago2}),
we have easily%
\begin{equation*}
\langle G_{\ell}(\alpha )\rangle =\sqrt{\ell+1}\left\{ \langle \widehat{F}_{\ell}(\alpha )\rangle -\alpha
\right\} \simeq \frac{1}{\sqrt{\ell+1}}b(\alpha )+o(\frac{1}{\ell})\text{ ,}
\end{equation*}%
so that the bias is negligible for $\ell$ large if one focuses on a single
row, whereas considering the whole array, i.e. summing from $\ell=1$ to $L,$
makes the bias nonnegligible in the aggregate.

The presence of estimated parameters has also a nonnegligible effects on the
behavior of the covariances, more precisely, it can be shown that we have,
as $L\rightarrow \infty ,$ and for all $0\leq \alpha _{1},\alpha _{2}\leq 1,$
$0\leq r_{1},r_{2}\leq 1,$%
\begin{eqnarray}
&&\lim_{L\rightarrow \infty }Cov\left\{ \widehat{K}_{L}(\alpha _{1},r_{1}),%
\widehat{K}_{L}(\alpha _{2},r_{2})\right\}  \notag \\
&=&\min (r_{1},r_{2})\min (\alpha _{1},\alpha _{2})\left\{ 1-\max (\alpha
_{1},\alpha _{2})\right\}  \notag \\
&&-\min (r_{1},r_{2})[(1-\alpha _{1})(1-\alpha _{2})\log (1-\alpha _{1})\log
(1-\alpha _{2})]\text{ ,}  \label{newcov}
\end{eqnarray}%
\ see Ref.\cite{mar} for further details.

In view of the previous results, for statistical inference we suggest to
focus on the mean-corrected field 
\begin{equation}
\label{eq:kfinal1}
\widehat{K}_{L}^{\prime }(\alpha ,r)=\widehat{K}_{L}(\alpha ,r)-2\sqrt{r}%
b(\alpha )\text{ .}
\end{equation}%
An alternative strategy would be to implement the bias correction directly on $G_\ell(\alpha)$, by using the exact (rather than asymptotic) expression for the bias which is provided in the appendix. Our experience with simulated maps suggests that the difference between these two approaches is quite negligible as a first approximation.
As before, it is possible to show a much stronger result than convergence of
the mean and variance, i.e., as $L\rightarrow \infty $%
\begin{equation*}
\widehat{K}_{L}^{\prime }(\alpha ,r)\Rightarrow \widehat{K}(\alpha ,r)\text{
,}
\end{equation*}%
where we write $\widehat{K}(\alpha ,r)$ for the zero mean Gaussian process
on $[0,1]\times \lbrack 0,1]$ with zero mean and covariance given in (\ref%
{newcov}); the convergence is meant in the uniform, functional sense
introduced before \cite{bill,dudley}. It is important
to notice that the limiting distribution does not depend on any nuisance
parameter, i.e. the asymptotic distribution is completely model-independent,
given Gaussianity. The structure of the limiting covariance measure may have
some interest for statistical application: for instance, it is well-known
from the theory of Gaussian fields that maxima will approximately occur in
the region where the variance (equations \ref{paris} and \ref{newcov}) of the field itself peaks \cite{adler}. In our
case, this corresponds roughly to $\alpha \simeq 0.4,$ $r\simeq 1$: a
maximum located far away from this value may by itself suggest some
non-Gaussianity in the $a_{\ell m}.$

The previous results, however, afford much more powerful and
well-established tests for Gaussianity. For instance, a Kolmogorov-Smirnov
type of test is implemented, for any suitably large $L,$ if we evaluate%
\begin{equation}
S_{L}^{(1)}=\sup_{0\leq r\leq 1}\sup_{0\leq \alpha \leq 1}|\widehat{K}%
_{L}^{\prime }(\alpha ,r)|\text{ ,}  \label{KS}
\end{equation}%
and compare the observed value with the threshold level obtained, for any
desired size of the test, by 
\begin{equation*}
S_{\infty }^{(1)}=\sup_{0\leq r\leq 1}\sup_{0\leq \alpha \leq 1}|\widehat{K}%
^{\prime }(\alpha ,r)|\text{ ;}
\end{equation*}%
the latter value can be readily derived by Monte Carlo simulation; note that
the limiting distribution does not entail any unknown, nuisance parameter.
Likewise, a Cramer-Von Mises type test can be implemented by looking at (see Ref.\cite{shorac}) 
\begin{equation}
S_{L}^{(2)}=\int_{0}^{1}\int_{0}^{1}(\widehat{K}_{L}^{\prime }(\alpha
,r))^{2}d\alpha dr\text{ .}  \label{CVM}
\end{equation}%
The relative performance of (\ref{KS}) and (\ref{CVM}) depends on the nature
of departure from Gaussianity; for instance, (\ref{KS}) may perform better in
cases when there is a relatively strong non-Gaussianity, concentrated on a
limited subsets of multipoles, whereas (\ref{CVM}) is better suited for
circumstances where the non-Gaussian behavior is more evenly spread over
many different angular frequencies. Likewise many other types of
goodness-of-fit statistics can be easily implemented, based upon the notion
that the field $\widehat{K}_{L}^{\prime }(\alpha ,r)$ should diverge, at
least for some $(\alpha ,r),$ if non-Gaussianity is truly present in the
marginal distribution of the spherical harmonic coefficients.

%

\section{The multivariate empirical process}
\label{sect:multivar}

The procedure of the previous Section is centered upon the search for
non-Gaussianity in the marginal distribution of the $\left\{ a_{\ell m}\right\} $%
. The empirical results presented in Section \ref{sect:empres} for some toy models show that
the resulting tests are very efficient indeed, if the $\left\{
a_{\ell m}\right\} $ are actually non-Gaussian. For some other toy models,
namely those where the non-Gaussianity is strong in pixel space, the
proposed procedure turns out to have much less power. This can be
intuitively explained as follows. Take $T(\theta ,\varphi )$ to be, for
instance, some nonlinear transform of an underlying Gaussian field, and
evaluate the spherical harmonic coefficients by (\ref{alm}); then, due to a
central limit theorem argument, the resulting linear combinations over the
pixels can have a distribution much closer to Gaussian in harmonic space.
 In this case the non-Gaussianity may show up as a dependency between $a_{\ell m}$ of different multipoles. With this in mind, in the present section we develop some more powerful
tests for non-Gaussianity, which consider the joint distribution of the $%
a_{\ell m}$ coefficients over different $\ell$ and $m$ values. More precisely, in the
bivariate case we look at the joint empirical distribution function%
\begin{eqnarray*}
\widehat{F}_{\ell_{1}\ell_{2}}(\alpha _{1},\alpha _{2}) &=&\frac{1}{\ell_{1}+1}%
\sum_{m_{1}=0}^{\ell_{1}}\underline{\mathbf{1}}(\widehat{u}_{\ell_{1}m}\leq \alpha _{1},\widehat{u}%
_{\ell_{2},m+\Delta _{m}}\leq \alpha _{2}) \\
&=&\frac{1}{\ell_{1}+1}\sum_{m_{1}=0}^{\ell_{1}}\underline{\mathbf{1}}(\widehat{u}_{\ell_{1}m}\leq \alpha
_{1})\underline{\mathbf{1}}(\widehat{u}_{\ell_{2},m+\Delta _{m}}\leq \alpha _{2})\text{ , }
\end{eqnarray*}%
for some integer $\Delta _{m}\geq 0$; in the sequel, we shall always use the
sum modulus $\ell_{2},$ e.g., $m+\Delta _{m}=m+\Delta _{m}-\left[\frac{m+\Delta_m}{\ell_2}\right]\ell_{2},$ if $m+\Delta
_{m}>\ell_{2}$ (here $[x]$ means integer-value) . In the general multivariate case we focus on%
\begin{equation}
\label{eq:li}
\widehat{F}_{\ell_{1}...\ell_{k}}(\alpha _{1},...,\alpha _{k})=\frac{1}{(\ell_{1}+1)}%
\sum_{m=0}^{\ell_{1}}\left\{ \underline{\mathbf{1}}(\widehat{u}_{\ell_{1}m}\leq \alpha
_{1})\prod_{i=2}^{k}\underline{\mathbf{1}}(\widehat{u}_{\ell_{i},m+\Delta _{mi}}\leq \alpha
_{i})\right\} \text{ , }\Delta _{mi}\geq 0\text{ .}
\end{equation}
Again if the $a_{\ell m}$ are Gaussian, $\widehat{F}_{\ell_{1}l_{2}}(\alpha _{1},\alpha _{2})\rightarrow\alpha_1\alpha_2$ (or $\widehat{F}_{\ell_{1}...\ell_{k}}(\alpha _{1},...,\alpha _{k})\rightarrow\prod_{i=1}^{k}\alpha _{i}$) and we obtain the multivariate empirical process in a similar way as for the univariate process%
\begin{equation*}
\widehat{G}_{\ell_{1}\ell_{2}}(\alpha _{1},\alpha _{2})=\sqrt{(\ell_{1}+1)}\left\{ 
\widehat{F}_{\ell_{1}\ell_{2}}(\alpha _{1},\alpha _{2})-\alpha _{1}\alpha
_{2}\right\} \text{ ,}
\end{equation*}%
and%
\begin{equation*}
\widehat{G}_{\ell_{1}...\ell_{k}}(\alpha _{1},...,\alpha _{k})=\sqrt{(\ell_{1}+1)}%
\left\{ \widehat{F}_{\ell_{1}...\ell_{k}}(\alpha _{1},...,\alpha
_{k})-\prod_{i=1}^{k}\alpha _{i}\right\} \text{ .}
\end{equation*}%
In the sequel, it is convenient to write $\ell_{1}=\ell$ and $\ell_{i+1}=\ell+\Delta
_{\ell i},$ for some positive and distinct integers $0<\Delta _{\ell 1},\Delta
_{\ell 2},...,\Delta _{\ell,k-1}<<L.$ For $0\leq \alpha _{1},\alpha _{2}\leq 1$, $%
0\leq r\leq 1$, we hence focus on 
\begin{equation*}
\widehat{K}_{L}(\alpha _{1},\alpha _{2},r)=\frac{1}{\sqrt{L-\Delta _{\ell 1}}}%
\sum_{\ell=1}^{[(L-\Delta _{\ell 1})r]}\widehat{G}_{\ell,\ell+\Delta _{\ell 1}}(\alpha
_{1},\alpha _{2})\text{ , }
\end{equation*}%
As discussed in Section \ref{sect:univar} using the estimated $C_\ell$ gives a bias which needs to be corrected for.
Some simple computations show that 
\begin{equation*}
\langle \prod_{i=1}^{k}\underline{\mathbf{1}}(\widehat{u}_{\ell_{i}m_{i}}\leq \alpha
_{i})\rangle =\prod_{i=1}^{k}\left\{ \alpha _{i}+\frac{1}{\ell}b(\alpha _{i})+o(\frac{1%
}{\ell})\right\}
\end{equation*}%
\begin{equation}
\label{eq:biasgen}
=\prod_{i=1}^{k}\alpha _{i}+\frac{1}{\ell}\sum_{i=1}^{k}\prod_{j=1,j\neq
i}^{k}\alpha _{j}b(\alpha _{i})+o(\frac{1}{\ell})\text{ .}
\end{equation}%
This gives the bias corrected $\widehat{K}_{L}^{\prime }(\alpha _{1},\alpha _{2},r)$
\begin{equation}
\label{eq:kfinal2}
\widehat{K}_{L}^{\prime }(\alpha _{1},\alpha _{2},r)=\widehat{K}_{L}(\alpha
_{1},\alpha _{2},r)-2\sqrt{r}\alpha _{1}b(\alpha _{2})-2\sqrt{r}\alpha
_{2}b(\alpha _{1})\text{ .}
\end{equation}
For the general multivariate case, we take the integers $\Delta _{\ell i}$ to be
strictly increasing, $0<\Delta _{\ell 1}<\Delta _{\ell 2}<...<\Delta _{\ell,k-1}<<L,$
and we consider%
\begin{equation*}
\widehat{K}_{L}(\alpha _{1},...,\alpha _{k},r)=\frac{1}{\sqrt{L-\Delta
_{\ell,k-1}}}\sum_{\ell=1}^{[(L-\Delta _{k-1})r]}\widehat{G}_{\ell,...,\ell+\Delta
_{\ell,k-1}}(\alpha _{1},...,\alpha _{k})\text{ ,}
\end{equation*}%
\begin{equation}
\label{eq:kfinal3}
\widehat{K}_{L}^{\prime }(\alpha _{1},...,\alpha _{k},r)=\widehat{K}%
_{L}(\alpha _{1},...,\alpha _{k},r)-2\sqrt{r}\sum_{i=1}^{k}\prod_{j=1,j\neq
i}^{k}\alpha _{j}b(\alpha _{i})\text{ .}
\end{equation}%
The rationale for these procedures can be explained as follows; although, as
motivated before, the marginal distribution of the spherical harmonic
coefficients can be close to Gaussian even in circumstances where $T(\theta
,\varphi )$ is not, the joint assumption that the $a_{\ell m}$ are Gaussian and
independent uniquely identifies a Gaussian field in pixel space, i.e., it
provides necessary and sufficient conditions. Therefore by looking at
multiple rows a non-Gaussian feature can be more likely detected.

By an argument similar to Section \ref{sect:univar}, it is possible to establish the weak
convergence of \ $\widehat{K}_{L}^{\prime }(\alpha _{1},...,\alpha _{k},r)$
to zero mean Gaussian fields with a complicated covariance function, which
is not reported for brevity's sake. We stress that the limiting distribution
changes when we vary the number of rows considered; on the other hand, for $%
L $ large the effect of the spacing factors $\Delta _{\ell},\Delta _{m}$ is
minimal (in the Gaussian case): hence, at least as a first approximation, a
single Monte Carlo tabulation is sufficient, given the number of rows
included. The robustness of the limiting result for varying $\Delta ^{\prime
}$s is further investigated in the empirical section.

In the previous discussion, we focussed for simplicity on the case where
only a single spherical harmonic coefficient was selected from any multipole 
$\ell$. This assumption may be relaxed to consider the joint distribution of
the spherical harmonic coefficients within the same row $\ell$. The limiting
form of the bias term is here more complicated, however, due to the
dependence among the normalized coefficients $\widehat{u}_{\ell m}$ over the
same row $\ell$. To evaluate this bias, we can provide the following general
result: for any $0\leq \alpha _{1},...,\alpha _{p}\leq 1$ such that 
\begin{equation*}
-\sum_{i=1}^{p}\frac{2\log (1-\alpha _{i})}{2\ell+1}\leq 1\text{ },
\end{equation*}%
(a condition which is always fulfilled, provided $\ell$ is large enough), we
have%
\begin{equation}
\langle \prod_{i=1}^{p}\underline{\mathbf{1}}(\widehat{u}_{\ell m_{i}}\leq \alpha _{i})\rangle =\sum_{\gamma \in
\Gamma _{p}}(-1)^{\#(\gamma )}\left( 1+\sum_{j\in \gamma }\frac{2\log
(1-\alpha _{j})}{2\ell+1}\right) ^{\ell-1/2},  \label{formsab}
\end{equation}
where $\Gamma _{p}$ is the class of all $2^{p}$ subsets $\gamma $ of $%
\left\{ 1,2,...,p\right\} ,$ and $\#(\gamma )$ denotes the number of
elements in $\gamma .$ For instance in the bivariate case $p=2$ we have $%
\Gamma _{p}=\left\{ \emptyset ,(1),(2),(1,2)\right\} $ and we obtain, for $\ell$
large enough,%
\begin{eqnarray*}
&\langle &\underline{\mathbf{1}}(\widehat{u}_{\ell m_{1}}\leq \alpha _{1})\underline{\mathbf{1}}(\widehat{u}_{\ell m_{2}}\leq \alpha
_{2})\rangle  \\
&=&1-\left( 1+\frac{2\log (1-\alpha _{1})}{2\ell+1}\right) ^{\ell-1/2}-\left( 1+%
\frac{2\log (1-\alpha _{2})}{2\ell+1}\right) ^{\ell-1/2} \\
&&+\left( 1+\frac{2\log (1-\alpha _{1})}{2\ell+1}+\frac{2\log (1-\alpha _{2})}{%
2\ell+1}\right) ^{\ell-1/2} \\
&=&\alpha _{1}\alpha _{2}+\frac{1}{\ell}\alpha _{2}b\left( \alpha _{1}\right) +%
\frac{1}{\ell}\alpha _{1}b(\alpha _{2})+\frac{1}{\ell}c(\alpha _{1},\alpha _{2})+o(%
\frac{1}{\ell})\text{ ,}
\end{eqnarray*}%
where we defined%
\begin{equation*}
c(\alpha _{1},\alpha _{2})=-\log \left( 1-\alpha _{1}\right) \log \left(
1-\alpha _{2}\right) (1-\alpha _{1})(1-\alpha _{2})\text{ .}
\end{equation*}%
Likewise, for $\ell$ large enough we have also%
\begin{eqnarray*}
&\langle&\underline{\mathbf{1}}(\widehat{u}_{\ell m_{1}}\leq \alpha _{1})\underline{\mathbf{1}}(\widehat{u}_{\ell m_{2}}\leq \alpha
_{2})\underline{\mathbf{1}}(\widehat{u}_{\ell m_{3}}\leq \alpha _{3})\rangle  \\
&=&1-\sum_{i=1}^{3}\left( 1+\frac{2\log (1-\alpha _{i})}{2\ell+1}\right)
^{\ell-1/2}+\sum_{i=1}^{3}\left( 1+\sum_{j=1,j\neq i}^{3}\frac{2\log (1-\alpha
_{j})}{2\ell+1}\right) ^{\ell-1/2} \\
&&-\left( 1+\sum_{j=1}^{3}\frac{2\log (1-\alpha _{j})}{2\ell+1}\right) ^{\ell-1/2}
\end{eqnarray*}%
\begin{eqnarray*}
&=&\alpha _{1}\alpha _{2}\alpha _{3}+\frac{1}{\ell}\alpha _{1}\alpha
_{2}b\left( \alpha _{3}\right) +\frac{1}{\ell}\alpha _{1}\alpha _{3}b(\alpha
_{2})+\frac{1}{\ell}\alpha _{2}\alpha _{3}b(\alpha _{3}) \\
&&+\frac{1}{\ell}\alpha _{1}c(\alpha _{2},\alpha _{3})+\frac{1}{\ell}\alpha
_{2}c(\alpha _{2},\alpha _{3})+\frac{1}{\ell}\alpha _{3}c(\alpha _{2},\alpha
_{3})\text{ .}
\end{eqnarray*}%
In higher dimensions, the bias structure is readily provided by analogous
computations, and we do not report it here for brevity's sake.

From the previous results, it is simple to derive the exact distribution of
the normalized (and squared) spherical harmonic coefficients, in the
presence of estimated parameters. We note first that%
\begin{equation*}
\langle \prod_{i=1}^{p}\underline{\mathbf{1}}(\widehat{u}_{\ell m_{i}}\leq \alpha _{i})\rangle \equiv P(\widehat{u}%
_{\ell m_{1}}\leq \alpha _{1},...,\widehat{u}_{\ell m_{p}}\leq \alpha _{p})\text{ ,}
\end{equation*}%
always, by the definition of the indicator function. Also, $|a_{\ell m}|^{2}/%
\widehat{C}_{\ell}\leq l+\frac{1}{2},$ by construction. Hence, for $0\leq
x_{1},...,x_{p}\leq l+\frac{1}{2},$ we obtain from (\ref{formsab})%
\begin{equation*}
P(\frac{|a_{\ell m_{1}}|^{2}}{\widehat{C}_{\ell}}\leq x_{1},...,\frac{%
|a_{\ell m_{p}}|^{2}}{\widehat{C}_{\ell}}\leq x_{p})=\sum_{\gamma \in \Gamma
_{p}}(-1)^{\#(\gamma )}\left( 1-\sum_{j\in \gamma }\frac{2x_{j}}{2\ell+1}%
\right) ^{\ell-1/2}.
\end{equation*}%
For instance, in the bivariate case, for $0\leq x_{1},x_{2}\leq l+\frac{1}{2}
$ we have%
\begin{equation*}
P(\frac{|a_{\ell m_{1}}|^{2}}{\widehat{C}_{\ell}}\leq x_{1},\frac{|a_{\ell m_{2}}|^{2}}{%
\widehat{C}_{\ell}}\leq x_{2})
\end{equation*}%
\begin{equation*}
=1-\left( 1-\frac{2x_{1}}{2\ell+1}\right) ^{\ell-1/2}-\left( 1-\frac{2x_{2}}{2\ell+1}%
\right) ^{\ell-1/2}+\left( 1-\frac{2x_{1}}{2\ell+1}-\frac{2x_{2}}{2\ell+1}\right)
^{\ell-1/2}.
\end{equation*}%
The previous expressions may have some independent interest when considering
exact statistical inference for CMB data (for instance, in likelihood
analysis).


\section{Empirical section}
\label{sect:empres}

In this section we will test our method for two classes of toy models. First we will use a non-Gaussianity generated in spherical harmonic space which should be easily detected as our method is based on the spherical harmonic coefficients. In the second class of test models we will generate the non-Gaussianity in pixel space . For these models we expect a drop in the detection level at least for the univariate test, as the transformation to spherical harmonic space may make the spherical harmonic coefficients more Gaussian using a central limit theorem argument.

To generate a sequence of non-Gaussian $a_{\ell m}$, we consider first the same
model as in Ref.\cite{kogut},i.e.%
\[
a_{\ell m}^{NG}=a_{\ell m}^{R}+ia_{\ell m}^{I}\mathrm{ ,}
\]%
where $a_{\ell m}^{R},$ $a_{\ell m}^{I}$ are independent $\chi _{\nu }^{2}$
variables, normalized to have zero mean and the same angular power spectrum
as the Gaussian part of the model.

Our field is then defined as (Model 1) 
\begin{eqnarray*}
T(\theta ,\varphi ) &=&\sum_{\ell=1}^{\infty
}\sum_{m=-\ell}^{\ell}a_{\ell m}Y_{\ell m}(\theta ,\varphi )\mathrm{ ,} \\
a_{\ell m} &=&(1-\beta )a_{\ell m}^{G}+\beta a_{\ell m}^{NG}\mathrm{ .}
\end{eqnarray*}%
Note that $\left\{ a_{\ell m}^{G}\right\} $ and $\left\{ a_{\ell m}^{NG}\right\} $
are independent and have by construction an identical angular power
spectrum; the percentage of non-Gaussianity in the model is then uniquely
determined as 
\begin{equation}
f_{NG}=\frac{\beta ^{2}}{(1-\beta )^{2}+\beta ^{2}}\mathrm{ .}  \label{nonga}
\end{equation}
Needless to say, this toy model is unphysical, but it is helpful to
illustrate some characteristics of our approach. Note the bispectrum $%
B_{\ell_{1}l_{2}\ell_{3}}^{m_{1}m_{2}m_{3}}=\langle a_{\ell_{1}m_{1}}a_{\ell_{2}m_{2}}a_{\ell_{3}m_{3}}\rangle 
$ here is identically zero, unless $m_{1}=m_{2}=m_{3}=0;$ hence the
selection rules implied by Wigner's 3j coefficients are not satisfied and
the model is (slightly) anisotropic. We do not view this as a major
difficulty, however, as the effect is minimal and this model is introduced
here for purely expository purposes.

\begin{figure}[tbp]
\begin{center}
\leavevmode
\epsfig {file=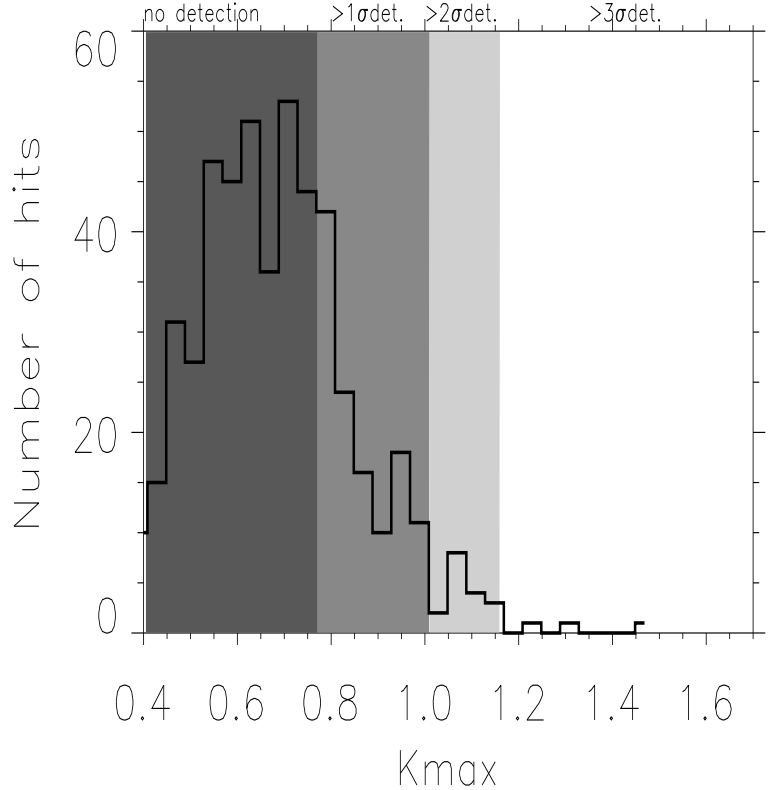,width=12cm,height=12cm}
\caption{The \protect{$K_\mathrm{max}$} (univariate) from a Monte Carlo simulation of 500 Gaussian CMB realizations. The colored areas indicate the \protect{$K_\mathrm{max}$} intervals where \protect{$68\%$}, \protect{$95\%$} and \protect{$99\%$} of the Gaussian realizations are found. If \protect{$K_\mathrm{max}$} is found in one of these intervals, this is taken as a \protect{$1$}, \protect{$2$} and \protect{$3$} sigma detection of non-Gaussianity respectively.}
\label{fig:mc1}
\end{center}
\end{figure}

\begin{figure}[tbp]
\begin{center}
\leavevmode
\epsfig {file=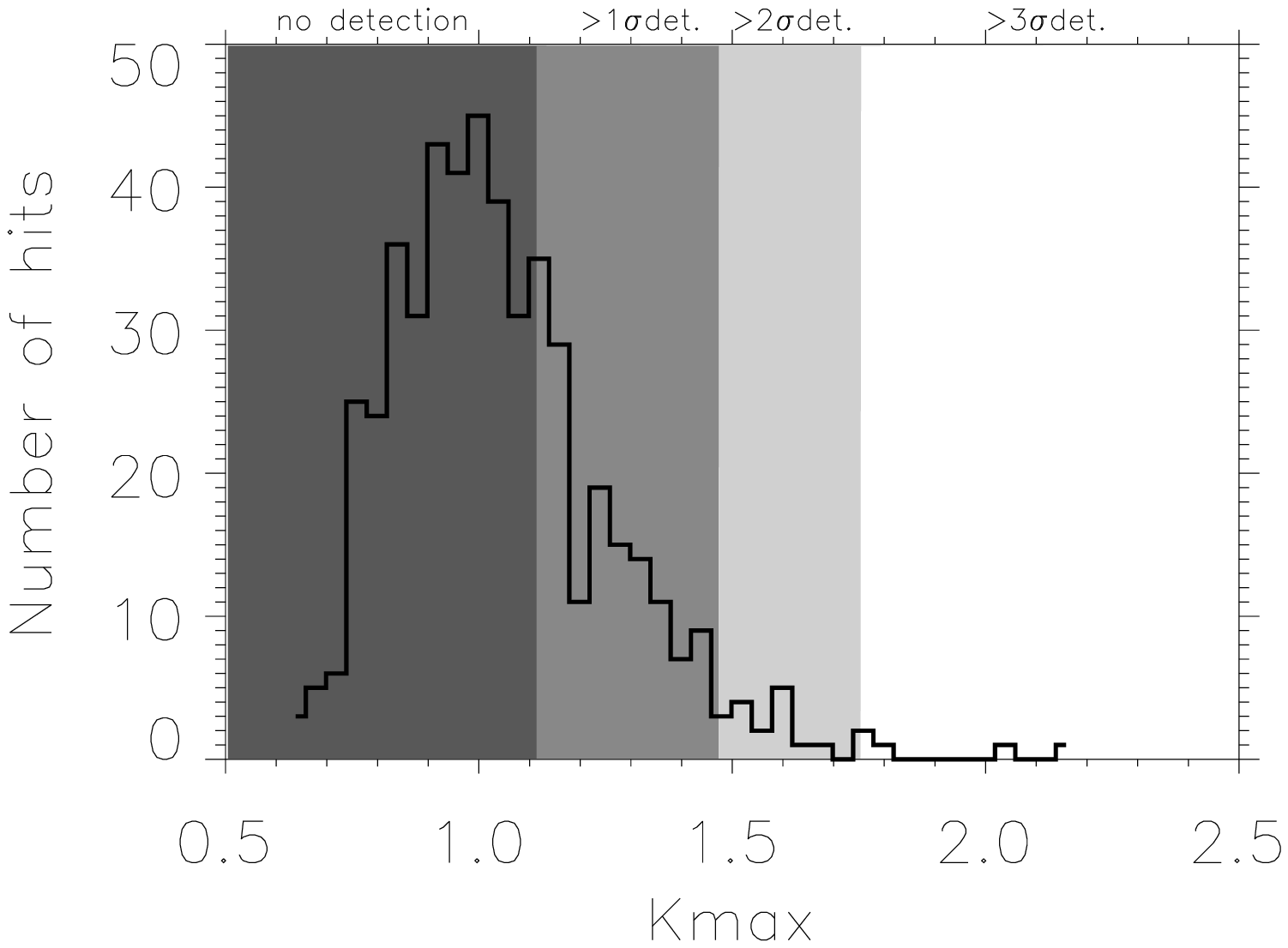,width=12cm,height=12cm}
\caption{The \protect{$K_\mathrm{max}$} (bivariate) from a Monte Carlo simulation of 500 Gaussian CMB realizations. The colored areas indicate the \protect{$K_\mathrm{max}$} intervals where $68\%$, $95\%$ and $99\%$ of the Gaussian realizations are found. If \protect{$K_\mathrm{max}$} is found in one of these intervals, this is taken as a $1$, $2$ and $3$ sigma detection of non-Gaussianity respectively.}
\label{fig:mc2}
\end{center}
\end{figure}

\begin{figure}[tbp]
\begin{center}
\leavevmode
\epsfig {file=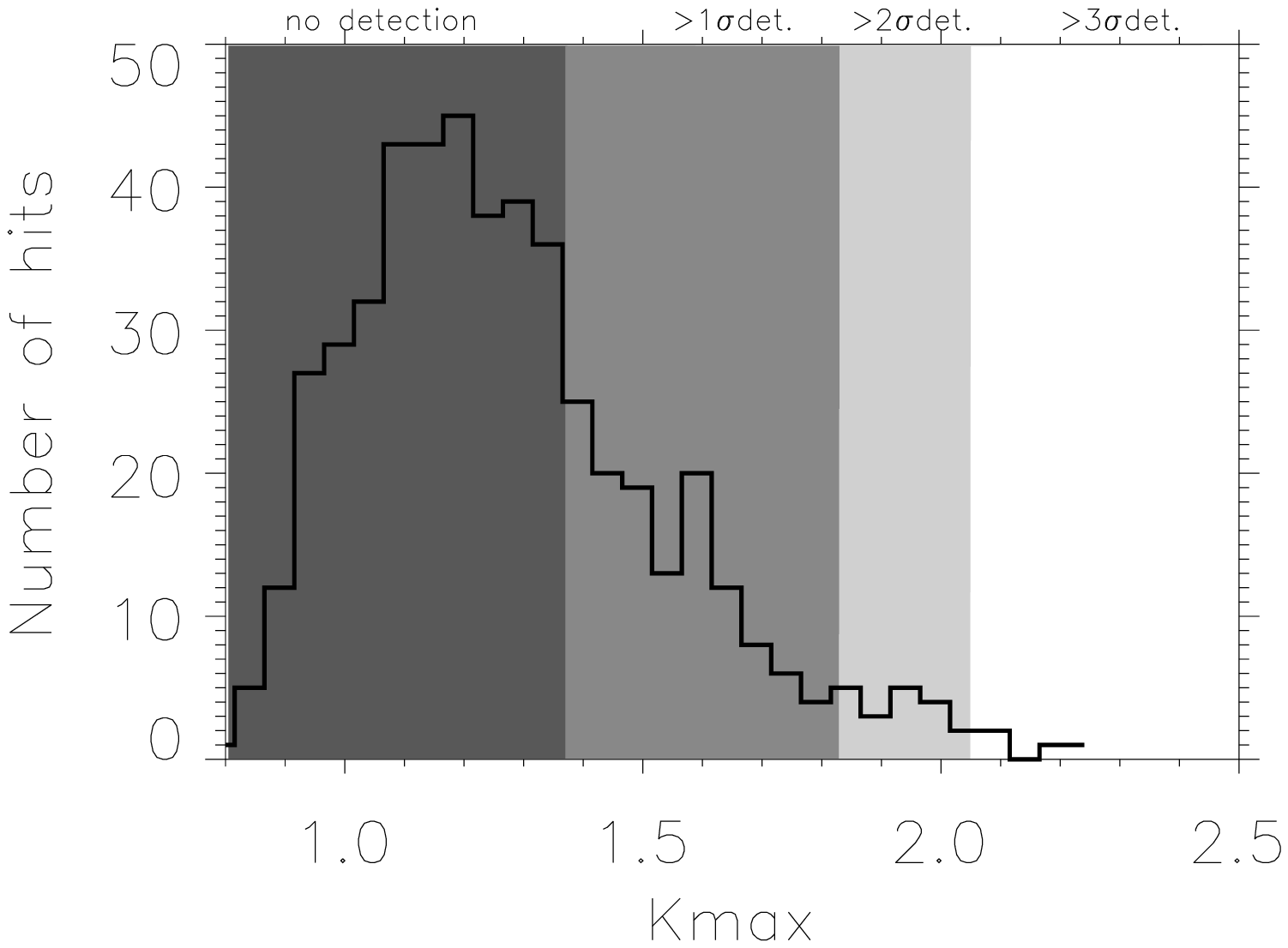,width=12cm,height=12cm}
\caption{The \protect{$K_\mathrm{max}$} (trivariate) from a Monte Carlo simulation of 100 Gaussian CMB realizations. The colored areas indicate the \protect{$K_\mathrm{max}$} intervals where $68\%$, $95\%$ and $99\%$ of the Gaussian realizations are found. If \protect{$K_\mathrm{max}$} is found in one of these intervals, this is taken as a $1$, $2$ and $3$ sigma detection of non-Gaussianity respectively.}
\label{fig:mc3}
\end{center}
\end{figure}

We consider Komogorov-Smirnov type tests as discussed in the previous sections, with a
straightforward extension to the multivariate case. We evaluate the distribution of $K_\mathrm{max}$ which is the sup of the $\hat K'_L(\alpha,r)$ function (see equations \ref{eq:kfinal1}, \ref{eq:kfinal2}, \ref{eq:kfinal3}) using Monte Carlo simulations of Gaussian CMB realizations. In figures (\ref{fig:mc1}), (\ref{fig:mc2}) and (\ref{fig:mc3}) we show the results for the uni-, bi- and trivariate case. We now define the $1\sigma$, $2\sigma$ and $3\sigma$ detection levels as the $K_\mathrm{max}$ values over which we had $68\%$, $95\%$ and $99\%$ of the hits in the Gaussian simulations respectively. These detection levels are shown as shades in the figures.
We consider the resolution $L=500;$ to approximate the suprema, we choose a
grid of 30 points over the $\alpha $-space, which makes the numerical
implementation of our Monte Carlo procedures extremely fast. We have some
evidence, however, that our results may be to some extent improved for finer
grids.

\begin{figure}[tbp]
\begin{center}
\leavevmode
\epsfig {file=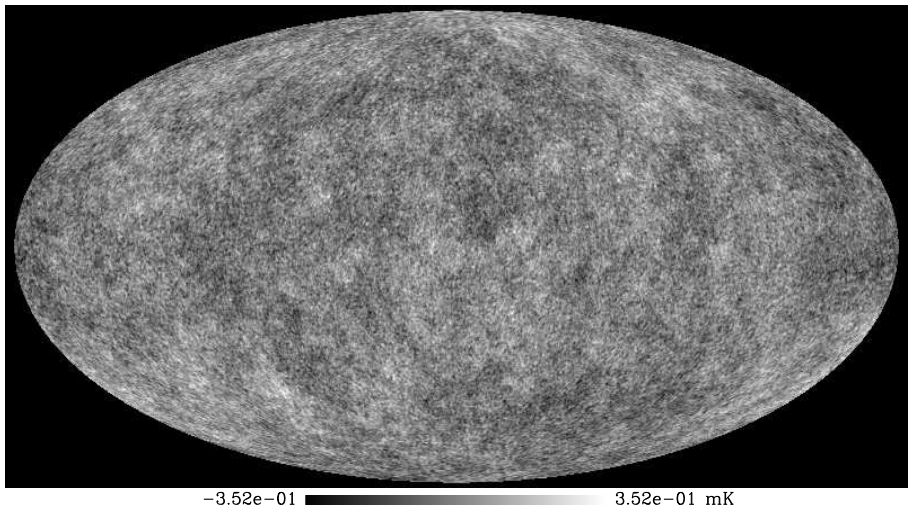,width=8cm,height=12cm,angle=90}
\caption{A Gaussian map (monopole and dipole removed) with a $5.9\%$ non-Gaussian part, made with $\chi^2$ distributed \protect{$a_\mathbf{\ell m}$}}
\label{fig:mapchi2}
\end{center}
\end{figure}

\begin{figure}[tbp]
\begin{center}
\leavevmode
\epsfig {file=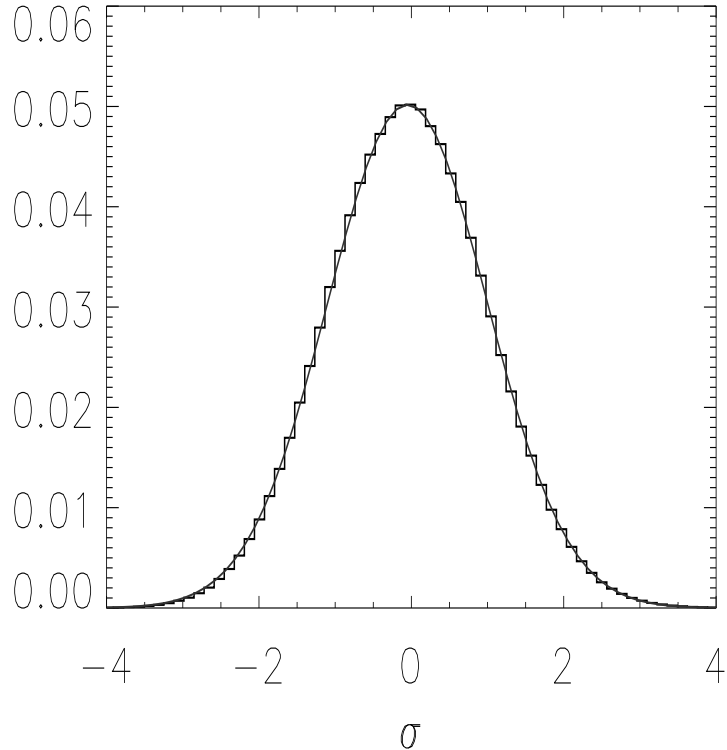,width=12cm,height=12cm}
\caption{The histogram of the pixels of a Gaussian map with a $5.9\%$ non-Gaussian part, made with $\chi^2$ distributed \protect{$a_\mathbf{\ell m}$}. The solid curve is a Gaussian fit to the histogram.}
\label{fig:histchi2}
\end{center}
\end{figure}

The results are reported in Tables \ref{tab:m1u}, \ref{tab:m1b} and \ref{tab:m1t}. We considered the univariate test
described in Section \ref{sect:univar}, and then the multivariate version of Section \ref{sect:multivar}; for
the latter, we focussed on bivariate and trivariate circumstances, with $%
\Delta \ell=250,$ $\Delta m=0$ and $\Delta \ell_{1}=\Delta m_{1}=249,$ $\Delta
\ell_{2}=\Delta m_{2}=250,$ respectively (see eq.\ref{eq:li}). 
For this model, the performance of the three
procedures is certainly very satisfactory. The power of the tests, i.e. the percentage of simulations where non-Gaussianity is correctly detected is a monotonic function of $\beta$ (the amount of non-Gaussianity) as expected. For instance for the univariate
test that can consistently detect as little as $6\%$ non-Gaussianity in the
map, at the $2\sigma$ level. A map of a model with $6\%$ non-Gaussian distributions are shown in figure (\ref{fig:mapchi2}) and its histogram in figure (\ref{fig:histchi2}). In pixel space this model seems to be almost identical to a Gaussian model.

\begin{figure}[tbp]
\begin{center}
\leavevmode
\epsfig {file=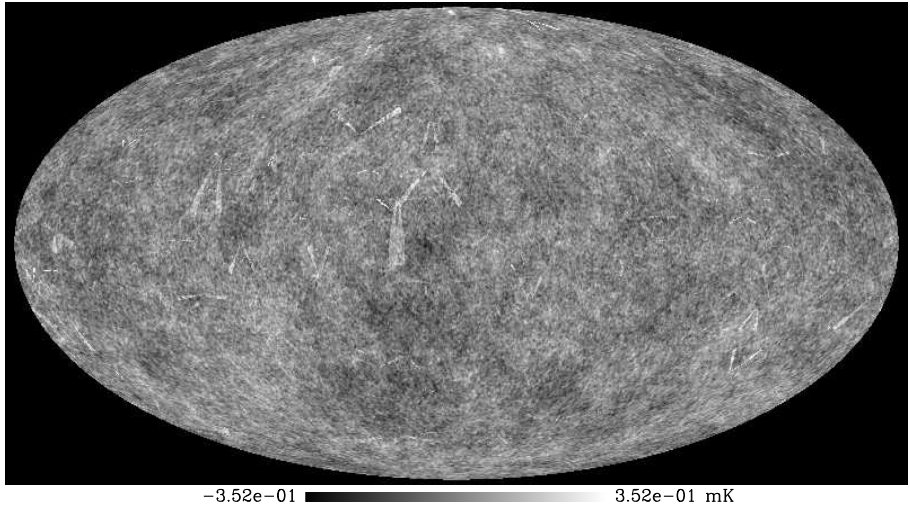,width=8cm,height=12cm,angle=90}
\caption{A Gaussian map (monopole and dipole removed) with a $15.5\%$ non-Gaussian part, made of string like objects of varying length and temperature. The map has a few pixels which have a much higher (positive) temperature than the rest of the map. These strong out-layers are here damped so that they don't dominate the map, making us unable to distinguish the other fluctuations. All pixels above \protect{$T_\mathrm{max}=|T_\mathrm{min}|$} are set to \protect{$T_\mathrm{max}$}.}
\label{fig:mapstrings}
\end{center}
\end{figure}

\begin{figure}[tbp]
\begin{center}
\leavevmode
\epsfig {file=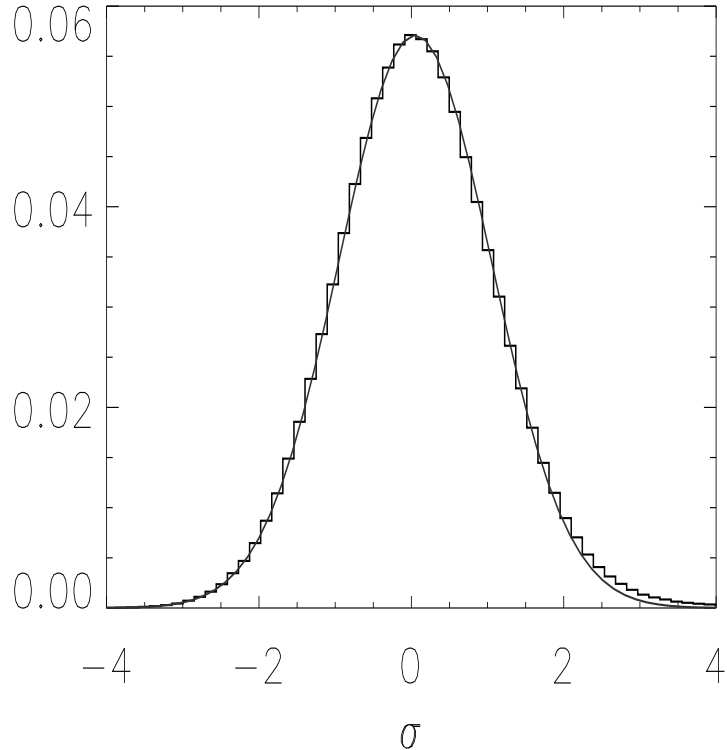,width=12cm,height=12cm}
\caption{The histogram of the pixels of the map showed in Figure (\protect{\ref{fig:mapstrings}}). This is a Gaussian map with a $15.5\%$ non-Gaussian part made of string like objects as described in the caption of Figure (\protect{\ref{fig:mapstrings}}). The solid curve is a Gaussian fit to the histogram. As small tail is seen due to the high temperature strings.}
\label{fig:histmapstrings}
\end{center}
\end{figure}

As a second alternative (Model 2), we try to mimic cosmic string models by
generating $10^{4}$ string-like features with randomly(Gaussian) varying length and temperature and superimposing them
on a Gaussian background, according to the expression%
\begin{equation}
T(\theta ,\varphi )=(1-\beta )T^G(\theta ,\varphi )+\beta T^{S}(\theta ,\varphi )\mathrm{ ,}
\label{strings}
\end{equation}%
where $T^{S}(\theta ,\varphi )$ is a pure string map, $T^G(\theta ,\varphi )$ is a Gaussian CMB map. The two fields are independent. The percentage of non-Gaussianity, $f_{NG}$, is again defined
by (\ref{nonga}). We view (\ref{strings}) as a model where non-Gaussianity
is strong in pixel space, and we thus expect our results not to be as good
as for the previous example. An inspection of the Tables \ref{tab:m2u}--\ref{tab:m2t2}, suggests that the
performance of the univariate test gives weaker results than what was the case for the non-Gaussianity generated in spherical harmonic space: at least $30\%$
of non-Gaussianity is needed to ensure a mere $50\%$ detections at $1\sigma$.
However, the
performance of the bivariate and trivariate procedures is more
promising, as non-Gaussianity can be detected for percentage of
non-Gaussianity around $10/15\%$ at the $1\sigma$ level. The effect of varying $\Delta m$ is
noticeable, but not extraordinary; likewise, some unreported simulations for
other values of $\Delta \ell$ (=$200$,$150$,$100$) show similar outcomes (on the
other hand, the performance is worse for very small $\Delta \ell$, i.e. $1$ to
$50$). In Figure (\ref{fig:mapstrings}) we show a map of a $15.5\%$ strings. The map is looking Gaussian except for a few strong out-layers which is seen as a tail in the histogram (Figure (\ref{fig:histmapstrings})). This map is at the limit of the $2\sigma$ detection level (for detection in $50\%$ of the cases) for our tests, but it seems that a test in pixel space may here give a stronger detection if the out-layers are not misinterpreted as foregrounds or noise.

\begin{center}
\begin{table}
\caption{MODEL 1, UNIVARIATE TEST}
\label{tab:m1u}
\begin{ruledtabular}
\begin{tabular}{|l|l|l|l|l|l|l|}
$\beta $ ($f_\mathrm{NG}$) & 0.20 (5.9\%) & 0.22 (7.3\%) & 0.24 (9\%) & 0.26 (11\%)
& 0.28 (13\%) & 0.30 (15.5\%) \\ 
\hline
1$\sigma $ & 78\% & 96\% & 100\% & 100\% & 100\% & 100\% \\ 
2$\sigma $ & 57\% & 82\% & 99\% & 100\% & 100\% & 100\% \\ 
3$\sigma $ & 39\% & 67\% & 95\% & 100\% & 100\% & 100\%
\end{tabular}
\end{ruledtabular}
\end{table}
\end{center}

\begin{center}
\begin{table}
\caption{MODEL 1, BIVARIATE TEST}
\label{tab:m1b}
\begin{ruledtabular}
\begin{tabular}{|l|l|l|l|l|l|l|}
$\beta $ ($f_\mathrm{NG})$ & 0.20 (5.9\%) & 0.22 (7.3\%) & 0.24 (9\%) & 0.26 (11\%)
& 0.28 (13\%) & 0.30 (15.5\%) \\
\hline 
1$\sigma $ & 61\% & 68\% & 90\% & 99\% & 100\% & 100\% \\ 
2$\sigma $ & 21\% & 31\% & 47\% & 82\% & 97\% & 100\% \\ 
3$\sigma $ & 6\% & 9\% & 19\% & 53\% & 90\% & 99\%
\end{tabular}
\end{ruledtabular}
\end{table}
\end{center}

\begin{center}
\begin{table}
\caption{MODEL 1, TRIVARIATE TEST}
\label{tab:m1t}
\begin{ruledtabular}
\begin{tabular}{|l|l|l|l|l|l|l|}
$\beta $ ($f_\mathrm{NG})$ & 0.20 (5.9\%) & 0.22 (7.3\%) & 0.24 (9\%) & 0.26 (11\%)
& 0.28 (13\%) & 0.30 (15.5\%) \\ 
\hline
1$\sigma $ & 32\% & 54\% & 76\% & 89\% & 96\% & 100\% \\ 
2$\sigma $ & 5\% & 17\% & 28\% & 67\% & 84\% & 98\% \\ 
3$\sigma $ & 1\% & 7\% & 12\% & 33\% & 67\% & 95\%
\end{tabular}
\end{ruledtabular}
\end{table}
\end{center}

\begin{center}
\begin{table}
\caption{MODEL 2, UNIVARIATE TEST}
\label{tab:m2u}
\begin{ruledtabular}
\begin{tabular}{|l|l|l|l|l|l|}
$\beta $ ($f_\mathrm{NG})$ & 0.20 (5.9\%)  & 0.25 (10\%) & 0.30 (15.5\%) & 0.40
(30.7\%) & 0.50 (50\%) \\ 
\hline
1$\sigma $ & 28\% & 33\% & 27\% & 48\% & 71\% \\ 
2$\sigma $ & 5\% & 5\% & 3\% & 17\% & 41\% \\ 
3$\sigma $ & 2\% & 2\% & 0\% & 3\% & 25\%
\end{tabular}
\end{ruledtabular}
\end{table}
\end{center}

\begin{center}
\begin{table}
\caption{MODEL 2, BIVARIATE TEST,$\Delta \ell=250,$ $\Delta m=0$ }
\label{tab:m2b}
\begin{ruledtabular}
\begin{tabular}{|l|l|l|l|l|l|}
$\beta $ ($f_\mathrm{NG})$ & 0.20 (5.9\%) & 0.25 (10\%) & 0.30 (15.5\%) & 0.40
(30.7\%) & 0.50 (50\%) \\
\hline 
1$\sigma $ & 33\% & 41\% & 55\% & 71\% & 83\% \\ 
2$\sigma $ & 1\% & 17\% & 22\% & 48\% & 57\% \\ 
3$\sigma $ & 0\% & 5\% & 9\% & 38\% & 44\%
\end{tabular}
\end{ruledtabular}
\end{table}
\end{center}

\begin{center}
\begin{table}
\caption{MODEL 2, TRIVARIATE TEST, $\Delta \ell_{1}=249,$ $\Delta \ell_{2}=250,$ $\Delta m_{1}=\Delta m_{2}=0$ }
\label{tab:m2t}
\begin{ruledtabular}
\begin{tabular}{|l|l|l|l|l|l|}
$\beta $ ($f_{NG})$ & 0.20 (5.9\%) & 0.25 (10\%) & 0.30 (15.5\%) & 0.40
(30.7\%) & 0.50 (50\%) \\ 
\hline
1$\sigma $ & 30\% & 31\% & 51\% & 72\% & 89\% \\ 
2$\sigma $ & 6\% & 17\% & 27\% & 56\% & 72\% \\ 
3$\sigma $ & 1\% & 5\% & 11\% & 46\% & 60\%
\end{tabular}
\end{ruledtabular}
\end{table}
\end{center}

\begin{center}
\begin{table}
\caption{MODEL 2, BIVARIATE TEST,$\Delta \ell=250,$ $\Delta m=250$ }
\label{tab:m2b2}
\begin{ruledtabular}
\begin{tabular}{|l|l|l|l|l|l|l|}
$\beta $ ($f_\mathrm{NG})$ & 0.20 (5.9\%) & 0.25 (10\%) & 0.30 (15.5\%) & 0.40
(30.7\%) & 0.50 (50\%) & 1.0 (100\%) \\
\hline 
1$\sigma $ & 46\% & 57\% & 73\% & 80\% & 96\% & 100\% \\ 
2$\sigma $ & 12\% & 27\% & 42\% & 65\% & 84\% & 98\% \\ 
3$\sigma $ & 4\% & 18\% & 26\% & 57\% & 77\% & 96\%%
\end{tabular}
\end{ruledtabular}
\end{table}
\end{center}

\begin{center}
\begin{table}
\caption{MODEL 2, TRIVARIATE TEST, $\Delta \ell_{1}=249,$ $\Delta \ell_{2}=250,$ $\Delta m_{1}=249$, $\Delta m_{2}=250$ }
\label{tab:m2t2}
\begin{ruledtabular}
\begin{tabular}{|l|l|l|l|l|l|l|}
$\beta $ ($f_\mathrm{NG})$ & 0.20 (5.9\%) & 0.25 (10\%) & 0.30 (15.5\%) & 0.40
(30.7\%) & 0.50 (50\%) & 1.0 (100\%) \\ 
\hline
1$\sigma $ & 39\% & 50\% & 65\% & 80\% & 90\% & 100\% \\ 
2$\sigma $ & 14\% & 27\% & 46\% & 72\% & 86\% & 100\% \\ 
3$\sigma $ & 6\% & 10\% & 32\% & 61\% & 81\% & 100\%
\end{tabular}
\end{ruledtabular}
\end{table}
\end{center}

\section{Comments and conclusions}
\label{sect:disc}

We believe the approach advocated here may enjoy some advantages over
existing methods, and we view it as complementary to geometric approaches in
pixel space (for instance, methods based an Minkowski functionals, local
curvature, or other topological properties). Our proposal allows for a
rigorous asymptotic theory; it is completely model free; it provides
information not only on the existence of non-Gaussianity, but also on its
location in the space of multipoles; the effect of estimated parameters is
carefully accounted for; given that the asymptotic behavior of the field $%
K_{L}$ has been thoroughly investigated, many other procedures, further than
those we considered here (Kolmogorov-Smirnov and Cramer-Von Mises tests) can
be immediately implemented; the extension to an arbitrary number of rows is
in principle straightforward (although computationally burdensome), whereas
for instance the explicit form of higher order cumulant spectra needs to be
derived analytically in a case by case fashion. Furthermore, our analysis of
the distributional properties of the spherical harmonic coefficients may
have some independent interest in other areas of CMB investigation.

Our approach is clearly related to the analysis of higher-order cumulant
spectra (such as the bispectrum and trispectrum) in harmonic space. Although
the bi- and trispectrum have been very widely used in empirical work, their
power properties against a variety of non-Gaussian models do not seem to
have been very much investigated. We conjecture that our procedure may enjoy
better power properties than higher order spectra in a number of
circumstances; heuristically, the bispectrum and the trispectrum search for
non-Gaussian features on the $a_{\ell m}$ by focusing essentially on their
skewness and kurtosis, whereas the method we advocate here probe their whole
multivariate distribution. The mutual interplay between these different
methodologies may lead to improvements in both directions: for instance, it
seems possible to model the bispectrum and trispectrum evaluated at
different multipoles as processes indexed by some $r$, much the same way as
we did here to combine the information from empirical processes $%
G_{\ell_{1}...\ell_{k}}$ into a single statistic $K_{L}.$ This may allow for a
more rigorous analysis in the aggregate, in order to understand whether a
single or a few high values are to be considered significant, over a set of $%
L$ statistics. On the other hand, incorporation of some selection rules
(e.g., the Wigner's $3j$ coefficients) into our analysis may help us to
exploit the isotropic nature of the field to probe non-Gaussianity more
efficiently. 

In this paper we have tested the method for two non-Gaussian models. One which was created in spherical harmonic space and the other which was created in pixel space. For the first model, we detect non--Gaussianity at a $2\sigma$ level 
(about $50\%$ of the times) with the univariate test even when the test model 
contains only $5.9\%$ non-Gaussianity. In this case the map is very similar to a Gaussian map. Our second non-Gaussian test model was generated in pixel space. In this case we had $2\sigma$ detections (about $50\%$ of the times) with $15.5\%$ non-Gaussianity. The map shows that this kind of non-Gaussianity may be more easily detected in pixel space. When taking into account realistic effects like detector noise and galactic cuts, the method has
so far shown promising results. This will be discussed further together with
tests on realistic non-Gaussian maps in \cite{paper2}.

\begin{acknowledgments}

We acknowledge the use of the Healpix package\footnote{http://www.eso.org/science/healpix/}.

\end{acknowledgments}

\bibliography{text+figs}

\begin{thebibliography}{32}
\expandafter\ifx\csname natexlab\endcsname\relax\def\natexlab#1{#1}\fi
\expandafter\ifx\csname bibnamefont\endcsname\relax
  \def\bibnamefont#1{#1}\fi
\expandafter\ifx\csname bibfnamefont\endcsname\relax
  \def\bibfnamefont#1{#1}\fi
\expandafter\ifx\csname citenamefont\endcsname\relax
  \def\citenamefont#1{#1}\fi
\expandafter\ifx\csname url\endcsname\relax
  \def\url#1{\texttt{#1}}\fi
\expandafter\ifx\csname urlprefix\endcsname\relax\def\urlprefix{URL }\fi
\providecommand{\bibinfo}[2]{#2}
\providecommand{\eprint}[2][]{\url{#2}}

\bibitem[{\citenamefont{Narlikar and Padbanabahn}(1991)}]{narlikar}
\bibinfo{author}{\bibfnamefont{J.~V.} \bibnamefont{Narlikar}} \bibnamefont{and}
  \bibinfo{author}{\bibfnamefont{T.}~\bibnamefont{Padbanabahn}},
  \bibinfo{journal}{ARAA} \textbf{\bibinfo{volume}{29}}, \bibinfo{pages}{325}
  (\bibinfo{year}{1991}).

\bibitem[{\citenamefont{Watson}()}]{watson}
\bibinfo{author}{\bibfnamefont{S.}~\bibnamefont{Watson}},
  \eprint{astro-ph/0005003}.

\bibitem[{\citenamefont{Martin et~al.}(2000)\citenamefont{Martin, Riazuelo, and
  Sakellariadou}}]{nongi1}
\bibinfo{author}{\bibfnamefont{J.}~\bibnamefont{Martin}},
  \bibinfo{author}{\bibfnamefont{A.}~\bibnamefont{Riazuelo}}, \bibnamefont{and}
  \bibinfo{author}{\bibfnamefont{M.}~\bibnamefont{Sakellariadou}},
  \bibinfo{journal}{Phys. Rev.} \textbf{\bibinfo{volume}{D61}},
  \bibinfo{pages}{083518} (\bibinfo{year}{2000}).

\bibitem[{\citenamefont{Contaldi et~al.}(2000)\citenamefont{Contaldi, Bean, and
  Magueijo}}]{nongi2}
\bibinfo{author}{\bibfnamefont{C.~R.} \bibnamefont{Contaldi}},
  \bibinfo{author}{\bibfnamefont{R.}~\bibnamefont{Bean}}, \bibnamefont{and}
  \bibinfo{author}{\bibfnamefont{J.}~\bibnamefont{Magueijo}},
  \bibinfo{journal}{Phys. Lett.} \textbf{\bibinfo{volume}{B468}},
  \bibinfo{pages}{189} (\bibinfo{year}{2000}).

\bibitem[{\citenamefont{Linde and Mukhanov}(1997)}]{nongi3}
\bibinfo{author}{\bibfnamefont{A.}~\bibnamefont{Linde}} \bibnamefont{and}
  \bibinfo{author}{\bibfnamefont{V.}~\bibnamefont{Mukhanov}},
  \bibinfo{journal}{Phys. Rev.} \textbf{\bibinfo{volume}{D56}},
  \bibinfo{pages}{535} (\bibinfo{year}{1997}).

\bibitem[{\citenamefont{Bartolo et~al.}()\citenamefont{Bartolo, Matarrese, and
  Riotto}}]{nongi4}
\bibinfo{author}{\bibfnamefont{N.}~\bibnamefont{Bartolo}},
  \bibinfo{author}{\bibfnamefont{S.}~\bibnamefont{Matarrese}},
  \bibnamefont{and} \bibinfo{author}{\bibfnamefont{A.}~\bibnamefont{Riotto}},
  \eprint{hep-ph/0112261}.

\bibitem[{\citenamefont{Gupta et~al.}()\citenamefont{Gupta, Berera, Heavens,
  and Matarrese}}]{nongi5}
\bibinfo{author}{\bibfnamefont{S.}~\bibnamefont{Gupta}},
  \bibinfo{author}{\bibfnamefont{A.}~\bibnamefont{Berera}},
  \bibinfo{author}{\bibfnamefont{A.~F.} \bibnamefont{Heavens}},
  \bibnamefont{and}
  \bibinfo{author}{\bibfnamefont{S.}~\bibnamefont{Matarrese}},
  \eprint{astro-ph/0205152}.

\bibitem[{\citenamefont{Gangui et~al.}({\natexlab{a}})\citenamefont{Gangui,
  Martin, and Sakellariadou}}]{nongi6}
\bibinfo{author}{\bibfnamefont{A.}~\bibnamefont{Gangui}},
  \bibinfo{author}{\bibfnamefont{J.}~\bibnamefont{Martin}}, \bibnamefont{and}
  \bibinfo{author}{\bibfnamefont{M.}~\bibnamefont{Sakellariadou}},
  \eprint{astro-ph/0205202}.

\bibitem[{\citenamefont{Gangui et~al.}({\natexlab{b}})\citenamefont{Gangui,
  Pogosian, and Winitzki}}]{strings}
\bibinfo{author}{\bibfnamefont{A.}~\bibnamefont{Gangui}},
  \bibinfo{author}{\bibfnamefont{L.}~\bibnamefont{Pogosian}}, \bibnamefont{and}
  \bibinfo{author}{\bibfnamefont{S.}~\bibnamefont{Winitzki}},
  \eprint{astro-ph/0112145}.

\bibitem[{\citenamefont{Novikov et~al.}(2000)\citenamefont{Novikov, Schmalzing,
  and Mukhanov}}]{novikov}
\bibinfo{author}{\bibfnamefont{D.}~\bibnamefont{Novikov}},
  \bibinfo{author}{\bibfnamefont{J.}~\bibnamefont{Schmalzing}},
  \bibnamefont{and} \bibinfo{author}{\bibfnamefont{V.~F.}
  \bibnamefont{Mukhanov}}, \bibinfo{journal}{A\&A}
  \textbf{\bibinfo{volume}{364}}, \bibinfo{pages}{17} (\bibinfo{year}{2000}).

\bibitem[{\citenamefont{Gott et~al.}(1990)}]{gott}
\bibinfo{author}{\bibfnamefont{J.~R.} \bibnamefont{Gott}} \bibnamefont{et~al.},
  \bibinfo{journal}{ApJ} \textbf{\bibinfo{volume}{352}}, \bibinfo{pages}{1}
  (\bibinfo{year}{1990}).

\bibitem[{\citenamefont{Eriksen et~al.}()\citenamefont{Eriksen, Banday, and
  G\'orski}}]{eriksen}
\bibinfo{author}{\bibfnamefont{H.~K.} \bibnamefont{Eriksen}},
  \bibinfo{author}{\bibfnamefont{A.~J.} \bibnamefont{Banday}},
  \bibnamefont{and} \bibinfo{author}{\bibfnamefont{K.~M.}
  \bibnamefont{G\'orski}}, \eprint{astro-ph/0206327}.

\bibitem[{\citenamefont{Heavens and Gupta}(2000)}]{heavens}
\bibinfo{author}{\bibfnamefont{A.~F.} \bibnamefont{Heavens}} \bibnamefont{and}
  \bibinfo{author}{\bibfnamefont{S.}~\bibnamefont{Gupta}},
  \bibinfo{journal}{MNRAS} \textbf{\bibinfo{volume}{324}}, \bibinfo{pages}{960}
  (\bibinfo{year}{2000}).

\bibitem[{\citenamefont{Scaramella and Vittorio}(1991)}]{vittorio}
\bibinfo{author}{\bibfnamefont{R.}~\bibnamefont{Scaramella}} \bibnamefont{and}
  \bibinfo{author}{\bibfnamefont{N.}~\bibnamefont{Vittorio}},
  \bibinfo{journal}{ApJ} \textbf{\bibinfo{volume}{375}}, \bibinfo{pages}{439}
  (\bibinfo{year}{1991}).

\bibitem[{\citenamefont{Dor\'e and S.~Colombi}()}]{dore}
\bibinfo{author}{\bibfnamefont{O.}~\bibnamefont{Dor\'e}} \bibnamefont{and}
  \bibinfo{author}{\bibfnamefont{F.~R.~B.} \bibnamefont{S.~Colombi}},
  \eprint{astro-ph/0202135}.

\bibitem[{\citenamefont{Phillips and Kogut}()}]{phillips}
\bibinfo{author}{\bibfnamefont{N.~G.} \bibnamefont{Phillips}} \bibnamefont{and}
  \bibinfo{author}{\bibfnamefont{A.}~\bibnamefont{Kogut}},
  \eprint{astro-ph/0112359}.

\bibitem[{\citenamefont{Komatsu et~al.}(2002)}]{komatsu}
\bibinfo{author}{\bibfnamefont{E.}~\bibnamefont{Komatsu}} \bibnamefont{et~al.},
  \bibinfo{journal}{ApJ} \textbf{\bibinfo{volume}{566}}, \bibinfo{pages}{19}
  (\bibinfo{year}{2002}).

\bibitem[{\citenamefont{Winitzki and Wu}()}]{win}
\bibinfo{author}{\bibfnamefont{S.}~\bibnamefont{Winitzki}} \bibnamefont{and}
  \bibinfo{author}{\bibfnamefont{J.~H.~P.} \bibnamefont{Wu}},
  \eprint{astro-ph/0007213}.

\bibitem[{\citenamefont{Hu}()}]{hu}
\bibinfo{author}{\bibfnamefont{W.}~\bibnamefont{Hu}},
  \eprint{astro-ph/0105117}.

\bibitem[{\citenamefont{Kunz et~al.}(2001)}]{kunz}
\bibinfo{author}{\bibfnamefont{M.}~\bibnamefont{Kunz}} \bibnamefont{et~al.},
  \bibinfo{journal}{ApJ} \textbf{\bibinfo{volume}{563}}, \bibinfo{pages}{L99}
  (\bibinfo{year}{2001}).

\bibitem[{\citenamefont{Wu et~al.}(2001)}]{boom}
\bibinfo{author}{\bibfnamefont{J.~H.} \bibnamefont{Wu}} \bibnamefont{et~al.},
  \bibinfo{journal}{Phys. Rev. Lett.} \textbf{\bibinfo{volume}{87}},
  \bibinfo{pages}{251303} (\bibinfo{year}{2001}).

\bibitem[{\citenamefont{Polenta et~al.}()}]{polenta}
\bibinfo{author}{\bibfnamefont{G.}~\bibnamefont{Polenta}} \bibnamefont{et~al.},
  \eprint{astro-ph/0201133}.

\bibitem[{\citenamefont{Ferreira et~al.}(1998)\citenamefont{Ferreira, Magueijo,
  and G\'orski}}]{cobeng1}
\bibinfo{author}{\bibfnamefont{P.~G.} \bibnamefont{Ferreira}},
  \bibinfo{author}{\bibfnamefont{J.}~\bibnamefont{Magueijo}}, \bibnamefont{and}
  \bibinfo{author}{\bibfnamefont{K.~M.} \bibnamefont{G\'orski}},
  \bibinfo{journal}{ApJ} \textbf{\bibinfo{volume}{503}}, \bibinfo{pages}{L1}
  (\bibinfo{year}{1998}).

\bibitem[{\citenamefont{Banday et~al.}(2000)\citenamefont{Banday, Zaroubi, and
  G\'orski}}]{cobeng2}
\bibinfo{author}{\bibfnamefont{A.~J.} \bibnamefont{Banday}},
  \bibinfo{author}{\bibfnamefont{S.}~\bibnamefont{Zaroubi}}, \bibnamefont{and}
  \bibinfo{author}{\bibfnamefont{K.~M.} \bibnamefont{G\'orski}},
  \bibinfo{journal}{ApJ} \textbf{\bibinfo{volume}{533}}, \bibinfo{pages}{575}
  (\bibinfo{year}{2000}).

\bibitem[{\citenamefont{Shorack and Wellner}(1986)}]{shorac}
\bibinfo{author}{\bibfnamefont{G.}~\bibnamefont{Shorack}} \bibnamefont{and}
  \bibinfo{author}{\bibfnamefont{J.}~\bibnamefont{Wellner}},
  \emph{\bibinfo{title}{Empirical Processes with Applications to Statistics,
  Wiley Series in Probability and Mathematical Statistics}}
  (\bibinfo{publisher}{J. Wiley, New York}, \bibinfo{year}{1986}).

\bibitem[{\citenamefont{Dudley}(1999)}]{dudley}
\bibinfo{author}{\bibfnamefont{R.~M.} \bibnamefont{Dudley}},
  \emph{\bibinfo{title}{Uniform Central Limit Theorems}}
  (\bibinfo{publisher}{Cambridge University Press, Cambridge},
  \bibinfo{year}{1999}).

\bibitem[{\citenamefont{Billingsley}(1968)}]{bill}
\bibinfo{author}{\bibfnamefont{P.}~\bibnamefont{Billingsley}},
  \emph{\bibinfo{title}{Convergence of Probability Measures}}
  (\bibinfo{publisher}{Wiley, New York}, \bibinfo{year}{1968}).

\bibitem[{\citenamefont{Marinucci and Piccioni}(2002)}]{mar}
\bibinfo{author}{\bibnamefont{Marinucci}} \bibnamefont{and}
  \bibinfo{author}{\bibnamefont{Piccioni}}, \bibinfo{journal}{submitted}
  (\bibinfo{year}{2002}).

\bibitem[{\citenamefont{Adler}(1990)}]{adler}
\bibinfo{author}{\bibfnamefont{A.~R.} \bibnamefont{Adler}},
  \emph{\bibinfo{title}{An introduction to continuity, extrema, and related
  topics for general Gaussian processes, Institute of Mathematical Statistics
  Lecture Notes--Monograph Series, 12}} (\bibinfo{publisher}{Institute of
  Mathematical Statistics, Hayward, CA}, \bibinfo{year}{1990}).

\bibitem[{\citenamefont{Kogut et~al.}(1996)}]{kogut}
\bibinfo{author}{\bibfnamefont{A.}~\bibnamefont{Kogut}} \bibnamefont{et~al.},
  \bibinfo{journal}{ApJL} \textbf{\bibinfo{volume}{464}}, \bibinfo{pages}{L29}
  (\bibinfo{year}{1996}).

\bibitem[{\citenamefont{Hansen et~al.}(2002)\citenamefont{Hansen, Marinucci,
  Natoli, and Vittorio}}]{paper2}
\bibinfo{author}{\bibfnamefont{F.~K.} \bibnamefont{Hansen}},
  \bibinfo{author}{\bibfnamefont{D.}~\bibnamefont{Marinucci}},
  \bibinfo{author}{\bibfnamefont{P.}~\bibnamefont{Natoli}}, \bibnamefont{and}
  \bibinfo{author}{\bibfnamefont{N.}~\bibnamefont{Vittorio}},
  \bibinfo{journal}{in preparation}  (\bibinfo{year}{2002}).

\bibitem[{\citenamefont{Johnson and Kotz}(1972)}]{kotz}
\bibinfo{author}{\bibfnamefont{N.~L.} \bibnamefont{Johnson}} \bibnamefont{and}
  \bibinfo{author}{\bibfnamefont{S.~J.} \bibnamefont{Kotz}},
  \emph{\bibinfo{title}{Distributions in statistics: continuous multivariate
  distributions, Wiley Series in Probability and Mathematical Statistics.}}
  (\bibinfo{publisher}{John Wiley \& Sons, Inc., New York-London-Sydney},
  \bibinfo{year}{1972}).

\end{thebibliography}

\appendix

\section{}
The purpose of this appendix is to derive analytically some of the results concerning the bias due to estimated parameters in our procedure; more details can be found in \cite{mar}. Heuristically, we are going to justify the appearance of 'extra' terms in equations (\ref{drago1}), (\ref{eq:biasgen}) and (\ref{eq:kfinal3}). To this aim, note first that
\begin{eqnarray*}
1\left( \widehat{u}_{\ell m}\leq \alpha \right) &\equiv &1\left( \frac{%
2|a_{\ell m}|^{2}}{\widehat{C}_{\ell}}\leq -2\log (1-\alpha )\right) \\
&=&1\left( \frac{2|a_{\ell m}|^{2}}{\sum_{m=-\ell}^{\ell}|a_{\ell m}|^{2}}\leq -\frac{%
2\log (1-\alpha )}{2\ell+1}\right) \text{ },
\end{eqnarray*}%
and 
\begin{equation}
\left( \frac{|a_{\ell 0}|^{2}}{\sum_{m=-\ell}^{\ell}|a_{\ell m}|^{2}},\frac{2|a_{\ell 1}|^{2}}{%
\sum_{m=-\ell}^{\ell}|a_{\ell m}|^{2}},...,\frac{2|a_{\ell m}|^{2}}{%
\sum_{m=-\ell}^{\ell}|a_{\ell m}|^{2}}\right) \overset{d}{=}D(\frac{1}{2},1,...,1),
\label{diri}
\end{equation}%
$\overset{d}{=}$ denoting equality in distribution and $D(\frac{1}{2}%
,1,...,1)$ a so-called Dirichlet distribution with parameters $(\frac{1}{2}%
,1,...,1)$ (see Ref.\cite{kotz}). 

In the sequel, we shall write
for brevity 
\begin{equation*}
\xi _{\ell 0}=\frac{|a_{\ell 0}|^{2}}{\sum_{m=-\ell}^{\ell}|a_{\ell m}|^{2}}\text{ , }\xi
_{\ell m}=\frac{2|a_{\ell m}|^{2}}{\sum_{k=-\ell}^{\ell}|a_{\ell m}|^{2}}\text{ .}
\end{equation*}%
and%
\begin{equation*}
\widetilde{\alpha }_{j\ell}\overset{def}{=}\frac{\Phi _{1}^{-1}(\alpha _{j})}{%
2\ell+1}\text{ , }\alpha _{j\ell}\overset{def}{=}\frac{\Phi _{2}^{-1}(\alpha _{1})%
}{2\ell+1}=\frac{-2\log (1-\alpha _{j})}{2\ell+1}\text{ , }j=1,2\text{ .}
\end{equation*}

\subsection{A general result on asymptotic bias}

We select a number $p\leq\ell$ out of $\ell$ elements $\left( \xi _{\ell 1},...,\xi _{\ell p}\right)$; of course the univariate, bivariate, and trivariate cases follow by setting $p$ equal to $1$, $2$ and $3$, respectively. From \cite{kotz} we know that 
\begin{equation*}
\left( \xi _{\ell 1},...,\xi _{\ell p}\right) \overset{d}{=}D(1,1,...,l-p+\frac{1}{2})
\end{equation*}
Note that
\begin{eqnarray*}
&&\langle \prod_{i=1}^{p}\underline{\mathbf{1}}(\widehat{u}_{\ell m_{i}}\leq \alpha
_{i})\rangle \\
&&=\langle \underline{\mathbf{1}}(\widehat{u}_{\ell m_1}\leq\alpha_1)\underline{\mathbf{1}}(\widehat{u}_{\ell m_2}\leq\alpha_2)...\underline{\mathbf{1}}(\widehat{u}_{\ell m_p}\leq\alpha_p)\rangle \\
&&=\langle \underline{\mathbf{1}}(\widehat{u}_{\ell m_1}\leq\alpha_1, \widehat{u}_{\ell m_2}\leq\alpha_2, ... , \widehat{u}_{\ell m_p}\leq\alpha_p)\rangle \\
&&=\langle \underline{\mathbf{1}}\left( \xi _{\ell 1}\leq \alpha _{1\ell},...,\xi _{\ell p}\leq \alpha _{p\ell}\right)\rangle \\
&&=P\left( \xi _{\ell 1}\leq \alpha _{1\ell},...,\xi _{\ell p}\leq \alpha _{p\ell}\right).
\end{eqnarray*}%

The technical result we need is then the following (see also \cite{mar}):\\
\textbf{Lemma A1 }Let $\Gamma _{p}$ be the class of all $2^{p}$ subsets $%
\gamma $ of $\left\{ 1,2,...,p\right\}$. 
For any $0\leq \alpha _{1\ell},...,\alpha _{p\ell}\leq 1$ such that $%
\sum_{i=1}^{p}\alpha _{i\ell}\leq 1,$ we have 
\begin{equation*}
P\left( \xi _{\ell 1}\leq \alpha _{1\ell},...,\xi _{\ell p}\leq \alpha _{p\ell}\right)
=\sum_{\gamma \in \Gamma _{p}}(-1)^{\#(\gamma )}\left( 1-\sum_{j\in \gamma
}\alpha _{j\ell}\right) ^{\ell-1/2},
\end{equation*}
where $\#(\gamma )$ denotes the number of elements of $\gamma .$

\ \newline
Some very simple special cases of Lemma A1, which are of direct interest for
the arguments below and can be checked by direct evaluation of the
corresponding integrals, are as follows: for $p=1,$ $\gamma =\left\{
\emptyset \right\} ,\left\{ 1\right\} $%
\begin{equation*}
P\left( \xi _{\ell1}\leq \alpha _{1\ell}\right) =1-(1-\alpha _{1\ell})^{\ell-1/2},
\end{equation*}%
and for $p=2,$ $\gamma =\left\{ \emptyset \right\} ,\left\{ 1\right\}
,\left\{ 2\right\} ,\left\{ 1,2\right\} $%
\begin{equation*}
P\left( \xi _{\ell1}\leq \alpha _{1\ell},\xi _{\ell2}\leq \alpha _{2\ell}\right)
=1-(1-\alpha _{1\ell})^{\ell-1/2}-(1-\alpha _{2\ell})^{\ell-1/2}+(1-\alpha _{1\ell}-\alpha
_{2\ell})^{\ell-1/2}.
\end{equation*}

\qquad \newline
\textbf{Proof of Lemma A1 }
We shall give the proof by induction. For $p=1,$ we have
immediately 
\begin{equation*}
\xi _{\ell1}\overset{d}{=}\beta (1,l-\frac{1}{2})\text{ ,}
\end{equation*}%
$\beta (a,b)$ denoting a Beta-distributed random variables with parameters $%
(a,b).$ Therefore, for $0\leq \alpha _{1\ell}\leq 1,$%
\begin{equation*}
P\left( \xi _{\ell1}\leq \alpha _{1\ell}\right) =\frac{\Gamma (\ell+1/2)}{\Gamma
(1)\Gamma (\ell-1/2)}\int_{0}^{\alpha _{1\ell}}(1-u)^{\ell-3/2}du=1-(1-\alpha
_{1\ell})^{\ell-1/2}.
\end{equation*}%
For clarity of exposition, we consider explicitly also $p=2,$ where we
obtain, taking into accounts that the conditional distribution of $\xi
_{\ell2}/(1-\xi _{\ell1})$ is $\beta (1,\ell-3/2),$%
\begin{eqnarray*}
P\left( \xi _{\ell1}\leq \alpha _{1\ell},\xi _{\ell 2}\leq \alpha _{2\ell}\right) &=&(\ell-%
\frac{1}{2})\int_{0}^{\alpha _{1\ell}}(1-x)^{\ell-3/2}\left[ 1-(1-\frac{\alpha
_{2\ell}}{1-x})^{\ell-3/2}\right] dx \\
&=&(\ell-\frac{1}{2})\int_{0}^{\alpha _{1\ell}}(1-x)^{\ell-3/2}dx-(\ell-\frac{1}{2}%
)\int_{0}^{\alpha _{1\ell}}(1-x-\alpha _{2\ell})^{\ell-3/2}dx \\
&=&1-(1-\alpha _{1\ell})^{\ell-1/2}-\left( 1-\alpha _{2\ell}\right) ^{\ell-1/2}+\left(
1-\alpha _{1\ell}-\alpha _{2\ell}\right) ^{\ell-1/2}.
\end{eqnarray*}%
For general $p>2,$ we have (see Ref.\cite{kotz}) 
\begin{eqnarray*}
&&P\left( \xi _{\ell 1}\leq \alpha _{1\ell},...,\xi _{\ell p}\leq \alpha _{p\ell}\right) \\
&=&\frac{\Gamma (\ell+1/2)}{\Gamma (\ell-p+1/2)}\int_{0}^{\alpha
_{1\ell}}...\int_{0}^{\alpha _{p\ell}}\left( 1-\sum_{i=1}^{p}u_{1}\right)
^{\ell-p-1/2}du_{1}...du_{p},
\end{eqnarray*}%
which becomes, after the change of variables $w_{i}=u_{i}/(1-u_{p}),$ $%
i=1,2,...,p-1$%
\begin{eqnarray}
&=&\frac{\Gamma (\ell+1/2)}{\Gamma (\ell-p+1/2)}\int_{0}^{\alpha
_{p\ell}}(1-u_{p})^{\ell-3/2}\int_{0}^{\alpha _{1\ell}/(1-u_{3})}...\int_{0}^{\alpha
_{p-1\ell}/(1-u_{3})}\left( 1-\sum_{i=1}^{p-1}w_{1}\right)
^{\ell-p+1/2}dw_{1}...dw_{p}  \notag \\
&=&\frac{\Gamma (\ell+1/2)}{\Gamma (\ell-1/2)}\int_{0}^{\alpha
_{p\ell}}(1-u_{p})^{\ell-3/2}P\left( \xi _{\ell 1}\leq \frac{\alpha _{1\ell}}{1-u_{3}}%
,...,\xi _{\ell p}\leq \frac{\alpha _{p\ell}}{1-u_{3}}\right) du_{3}  \notag \\
&=&\frac{\Gamma (\ell+1/2)}{\Gamma (\ell-1/2)}\int_{0}^{\alpha
_{p\ell}}(1-u_{p})^{\ell-3/2}\left\{ \sum_{\gamma ^{\prime }\in \Gamma
_{p-1}}(-1)^{\#(\gamma )}\left( 1-\sum_{j\in \gamma }\frac{\alpha _{j\ell}}{%
1-u_{3}}\right) ^{\ell-3/2}\right\} du_{3},  \label{aspi}
\end{eqnarray}%
the last equality following from the induction step. Now (\ref{aspi}) becomes%
\begin{eqnarray*}
&&\sum_{\gamma ^{\prime }\in \Gamma _{p-1}}(-1)^{\#(\gamma )}\left\{ \frac{%
\Gamma (\ell+1/2)}{\Gamma (\ell-1/2)}\int_{0}^{\alpha _{p\ell}}\left(
1-u_{3}-\sum_{j\in \gamma ^{\prime }}\alpha _{j\ell}\right)
^{\ell-3/2}du_{3}\right\} \\
&=&\sum_{\gamma ^{\prime }\in \Gamma _{p-1}}(-1)^{\#(\gamma )}\left\{ \left(
1-\sum_{j\in \gamma ^{\prime }}\alpha _{j\ell}\right) ^{\ell-1/2}-\left( 1-\alpha
_{p\ell}-\sum_{j\in \gamma ^{\prime }}\alpha _{j\ell}\right) ^{\ell-1/2}\right\} \\
&=&\sum_{\gamma \in \Gamma _{p}}(-1)^{\#(\gamma )}\left( 1-\sum_{j\in \gamma
}\alpha _{j\ell}\right) ^{\ell-1/2},
\end{eqnarray*}%
as required, by posing $\gamma =(\gamma ^{\prime }\cup \left\{ p\right\} )$,
and because $\Gamma _{p}=\Gamma _{p-1}\cup \left\{ (\gamma ^{\prime
},p):\gamma ^{\prime }\in \Gamma _{p-1}\right\} .$

\ \hfill $\square $

\ \newline
From Lemma A1, and for $\ell$ suitably large%
\begin{eqnarray*}
\langle 1\left( \xi _{\ell m}\leq \frac{-2\log (1-\alpha )}{2\ell+1}\right)\rangle  &=&P(\xi
_{\ell m}\leq \frac{-2\log (1-\alpha )}{2\ell+1}) \\
&=&1-(1-\frac{(-2\log (1-\alpha )}{2\ell+1})^{\ell -1/2}\text{ , }\alpha \in
\lbrack 0,1]\text{ ,}
\end{eqnarray*}%
and it can be readily checked that%
\begin{equation*}
\lim_{\ell\rightarrow \infty }l\left\{ 1-(1+\frac{\log (1-\alpha )}{\ell+1/2}%
)^{\ell-1/2}-\alpha \right\} =b(\alpha )\text{ .}
\end{equation*}%
Finally, note that, as $L\rightarrow \infty $ 
\begin{equation*}
\langle K_{L}(x,r)\rangle \simeq \frac{b(\alpha )}{\sqrt{L}}\sum_{\ell=1}^{[Lr]}\frac{1}{\sqrt{%
\ell+1}}\simeq b(\alpha )\sqrt{r}\int_{0}^{1}\frac{1}{\sqrt{u}}du=2b(\alpha )%
\sqrt{r}\text{ .}
\end{equation*}%
The calculations for the multidimensional cases follow in very much the same
way.

\subsection{The bias for the $m=0$ term}

We show now that the bias for the term corresponding to $m=0$ is
asymptotically negligible. Note first that $\xi _{\ell 0}\overset{d}{=}\beta (%
\frac{1}{2},\ell)$, and define 
\begin{equation*}
T_{\ell}=\frac{\ell|a_{\ell 0}|^{2}}{\sum_{m=1}^{\ell}|a_{\ell m}|^{2}},
\end{equation*}%
which has a Snedecor $F$-distribution, with 1 and $2\ell$ degrees of freedom.
We have%
\begin{equation*}
\xi _{\ell 0}=\frac{|a_{\ell 0}|^{2}}{|a_{\ell 0}|^{2}+2\sum_{m=1}^{\ell}|a_{\ell m}|^{2}}=%
\frac{T_{\ell}}{2\ell+T_{\ell}},
\end{equation*}%
whence, after straightforward calculations
\begin{eqnarray}
&\langle &\underline{\mathbf{1}}\left( \Phi _{1}(\frac{|a_{\ell 0}|^{2}}{|a_{\ell 0}|^{2}+2%
\sum_{m=1}^{\ell}|a_{\ell m}|^{2}})\leq \alpha\right) \rangle -\alpha\\
&&=P\left( \Phi _{1}(\frac{|a_{\ell 0}|^{2}}{|a_{\ell 0}|^{2}+2%
\sum_{m=1}^{\ell}|a_{\ell m}|^{2}})\leq \alpha \right) -\alpha\\
&&=P(T_{\ell}\leq \frac{
2\ell\Phi _{1}^{-1}(\alpha )}{2\ell+1-\Phi _{1}^{-1}(\alpha )})-\alpha\\
&&=P(T_{\ell}\leq \frac{2\ell\Phi _{1}^{-1}(\alpha )}{2\ell+1-\Phi _{1}^{-1}(\alpha )}%
)-P(T_{\ell}\leq \Phi _{1}^{-1}(\alpha ))+P(T_{\ell}\leq \Phi _{1}^{-1}(\alpha
))-\alpha  \notag \\
\text{ } &&=P\left( T_{\ell}\leq \frac{\Phi _{1}^{-1}(\alpha )}{1+(1-\Phi
_{1}^{-1}(\alpha ))/2\ell}\right) -P(T_{\ell}\leq \Phi _{1}^{-1}(\alpha ))
\label{biaunob} \\
&&+P\left( T_{\ell}\leq \Phi _{1}^{-1}(\alpha )\right) -\alpha \text{ .}
\label{biadue}
\end{eqnarray}%
The first component (\ref{biaunob}) for $\ell$ large enough is bounded by%
\begin{eqnarray}
&&P\left( \frac{\Phi _{1}^{-1}(\alpha )}{1+|1-\Phi _{1}^{-1}(\alpha )|/2\ell}%
\leq T_{\ell}\leq \frac{\Phi _{1}^{-1}(\alpha )}{1-|1-\Phi _{1}^{-1}(\alpha
)|/2\ell}\right)  \notag \\
&=&P\left( \frac{\sqrt{\Phi _{1}^{-1}(\alpha )}}{\sqrt{1+|1-\Phi
_{1}^{-1}(\alpha )|/2\ell}}\leq \sqrt{T_{\ell}}\leq \frac{\sqrt{\Phi
_{1}^{-1}(\alpha )}}{\sqrt{1-|1-\Phi _{1}^{-1}(\alpha )|/2\ell}}\right)
\label{student} \\
&=&O(\frac{1}{\ell})\text{ ,}  \notag
\end{eqnarray}%
because $\sqrt{T_{\ell}}$ has a Student's $t$-distribution, the latter has a
bounded density function and the length of the interval in (\ref{student})
is of order $\ell^{-1}.$ For (\ref{biadue}), we have that%
\begin{equation}
P\left( T_{\ell}\leq \Phi _{1}^{-1}(\alpha )\right) -\alpha =2\left\{
\int_{0}^{z_{\alpha }}f_{\ell-1}(u)du-\int_{0}^{z_{\alpha }}\varphi
(u)du\right\} \text{ ,}  \label{tint}
\end{equation}%
$f_{\ell}(x)$ and $\varphi (x)$ denoting, respectively, the density functions
of a Student's $t$ distribution with $2\ell$ degrees of freedom, i.e. 
\begin{equation*}
f_{\ell}(u)=\frac{\Gamma ((\ell+1)/2)}{\Gamma (\ell/2)}\frac{1}{\sqrt{\ell\pi }}\frac{1}{%
(1+u^{2}/\ell)^{(\ell+1)/2}}\text{ ,}
\end{equation*}%
and a standard Gaussian density. Thus (\ref{tint}) is bounded by%
\begin{eqnarray}
&=&2\left\{ \frac{\Gamma ((\ell+1)/2)}{\sqrt{\ell/2}\Gamma (\ell/2)}\frac{1}{\sqrt{%
2\pi }}-\frac{1}{\sqrt{2\pi }}\right\} \int_{0}^{z_{\alpha }}\exp
(-u^{2}/2)du  \label{un} \\
&&+2\frac{\Gamma ((\ell+1)/2)}{\Gamma (\ell/2)}\frac{1}{\sqrt{\ell\pi }}%
\int_{0}^{z_{\alpha }}\left\{ (1+u^{2}/\ell)^{-\ell/2}-\exp (-u^{2}/2)\right\} du%
\text{ .}  \label{treso}
\end{eqnarray}%
For (\ref{treso}) we use 
\begin{equation*}
\sup_{0\leq u\leq x}\left[ \left( 1+\frac{u^{2}}{\ell}\right) ^{-(\ell+1)/2}-\exp
(-u^{2}/2)\right] =O(\frac{1}{\ell})\text{ },
\end{equation*}%
and finally, (\ref{un}) follows from the well-known property%
\begin{equation*}
\left\{ \frac{\Gamma ((\ell+1)/2)}{\sqrt{\ell/2}\Gamma (\ell/2)}-1\right\} =O(\frac{1%
}{\ell})\text{ , as }l\rightarrow \infty \text{ .}
\end{equation*}%
Thus it follows that (\ref{un}) is $O(1/\ell)$ as well, and the proof is
completed. \ \newline

\end{document}